\documentclass[aps,preprint,superscriptaddress, showpacs, showkeys]{revtex4}

\usepackage{graphicx}

\newcommand{\be}{\begin{equation}}
\newcommand{\ee}{\end{equation}}
\newcommand{\bea}{\begin{eqnarray}}
\newcommand{\eea}{\end{eqnarray}}

\newcommand{\vfi}{\varphi}
\newcommand{\Vm}{V^-}

\newcommand{\Va}{V_a}
\newcommand{\Vb}{V_b}

\newcommand{\tf}{\tilde{f}}
\newcommand{\cV}{{\cal V}}
\newcommand{\bal}{\bar{l}}

\newcommand{\LL}{\frac{1}{2\pi}\,\int_{-\infty}^{\infty}dk\sum_{n=-\infty}^{\infty}}
\newcommand{\LLnof}{\int_{-\infty}^{\infty}dk\sum_{n=-\infty}^{\infty}}
\newcommand{\FF}{\frac{1}{2\pi}\,\int\limits_{0}^{2\pi}\int\limits_{-\infty}^{\infty}d\vfi dz}

\begin{document}

\title{\vspace{0cm}\large\bf
Electrostatic Patch Effect in Cylindrical Geometry. II. Forces}
\author{Valerio Ferroni}
\affiliation{ICRANet, Dept. of Phys., Univ.  `La Sapienza', Rome, Italy\\
{\rm current address}: W.W.Hansen Experimental Physics Laboratory,\\
Stanford University, Stanford, CA 94305-4085, USA}
\email{vferroni@stanford.edu}
\author{Alexander S. Silbergleit}
\affiliation{Gravity Probe B, W.W.Hansen Experimental Physics Laboratory,\\
Stanford University, Stanford, CA 94305-4085, USA}
\email{gleit@stanford.edu}

\date{\today}

\begin{abstract}  
We continue our study of patch effect (PE) for two close cylindrical conductors with parallel axes, slightly shifted against each other in the radial and by any length in the axial direction, started in~\cite{VA}, where the potential and energy in the gap were calculated to the second order in the small transverse shift, and to lowest order in the gap to cylinder radius ratio. Based on these results, here we derive and analyze PE force. It consists of three parts: the usual capacitor force due to the uniform potential difference, the one from the interaction between the voltage  patches and the uniform voltage difference, and the force due to patch interaction, entirely independent of the uniform voltage. General formulas for these forces are found, and their general properties are described. A convenient model of a localized patch is then suggested that allows us to calculate all the forces in a closed elementary form. Using this, a detailed analysis of the patch interaction for one pair of patches is carried out, and the dependence of forces on the patch parameters (width and strength) and their mutual position is examined. We also give various estimates of the axial patch effect force important for the Satellite Test of the Equivalence Principle (STEP), and recommend intensive pre-flight simulations employing our results. 
\end{abstract}

\keywords{Electrostatics - Patch effect - Cylindrical capacitor - Forces - Precision measurements - STEP}
\pacs{41.20Cv; 02.30Em; 02.30Jr; 04.80Cc }

\maketitle

\section{Introduction\label{s1}}

Electrostatic patch effect~\cite{Dar} is a nonuniform potential distribution on the surface of a metal first examined theoretically in paper~\cite{Sp}, where the PE force in a plane capacitor with varying boundary voltages has been calculated. This analysis was particularly motivated by the LISA space experiment to detect gravitational waves (see c.f.~\cite{LISA}). PE can similarly affect the accuracy of any other precision measurement if its set-up includes conducting surfaces in a closed proximity to each other. PE torques turned out one of the two major difficulties~\cite{E, H&S, BT} in the analysis of data from Gravity Probe B (GP-B) Relativity Science Mission (2004--2005 flight), that measured relativistic drift of a gyroscope predicted by Einstein's general relativity ~\cite{KetAl}. This required theoretical calculation of PE torques~\cite{KKS} for the case of two concentric spherical conductors.

Here we continue studying PE in cylindrical geometry that we started in paper~\cite{VA}, henceforth referred to as CPEI, for "`Cylindrical Patch Effect"'. It is strongly motivated by the experimental configuration of STEP~\cite{PW, Mes, PWMes, Over}, where each test mass and its superconducting magnetic bearing is a pair of approximately coaxial conducting cylinders (see section \ref{s6.1} for more details). The goal of STEP is the precise (1 part in $10^{18}$) measurement of the relative axial acceleration of a pair of coaxial test masses, so the importance of properly accounting for PE forces is evident. Notably, the axial and transverse force we find is inversely proportional to the gap and its square, respectively, exactly as in the plane capacitor~\cite{Sp}.

We determine the PE forces between the two slightly shifted cylinders with parallel axes by the energy conservation argument: a small shift, $\vec{r}{\;^0}$, of one of the conductors relative to the other causes an electrostatic force given by (see, for instance,~\cite{Smy})
\be
\vec F(\vec{r}^{\;0}) = -\frac{\partial W(\vec{r}^{\;0})}{\partial\vec{r}^{\;0}},\qquad
\vec F^{\;0}(0) = -\;\frac{\partial W(\vec{r}{\;^0)}}{\partial\vec{r}^{\;0}}\Biggl|_{\;\vec{r}^{\;0}=0}\; ,
\label{forcegen}
\ee
where $W(\vec{r}{\;^0})$ is the electrostatic energy as a function of the shift. The latter was found in CPEI to the second order in the small transverse shift, $\rho_0\ll d$, where $d$ is the gap between the cylinders in the coaxial position, so the force to the first order in $\rho_0$ is found below. For typical experimental conditions, such as the STEP configuration ~\cite{Mes, Over}, the gap is much smaller than either of the cylinders' radii; thus two small parameters are actually involved in the problem, $\rho_0/d\,\ll\,1,\;\;  d/a\,\sim \,d/b\,\ll\,1$. While justifying the model of infinite cylinders, this also allows for a significant simplification of the results to l. o. in $d/a$.

In the next section we summarize the results from CPEI needed for the force calculation. Based on this, we calculate PE forces in section \ref{s3}, and illucidate their general properties. In section \ref{s4} we introduce a convenient model of a localized patch potential allowing one to find simple closed--form expressions for the forces. Section \ref{s5} contains a detailed analysis of PE forces when a single localized patch described by our model is sitting at each of the cylinder boundaries. In section \ref{s6} estimates of the axial patch force for the STEP experiment set--up are given. The details of calculations, in places rather complicated and cumbersome, are found in the appendix.

\section{Summary of Results from CPEI\label{s2}}

\begin{figure}[ht]
\centering
\includegraphics[scale=0.5]{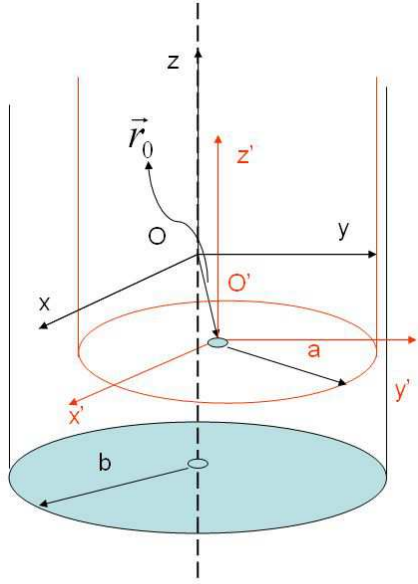}
\caption{{Geometry of the problem and coordinate systems}}
\label{fig1}
\end{figure}

We use Cartesian and cylindrical coordinates in two frames of the inner and outer cylinders as shown in fig. \ref{fig1}. In the inner, or `primed', frame the position of a point is given by the vector radius $\vec{r}\;^{'}$, and Cartesian coordinates $\{x^{\;'},y^{\;'},z^{\;'}\}$ or cylindrical coordinates $\{\rho^{\;'},\vfi^{\;'},z^{\;'}\}$. In the outer, or `unprimed', frame the corresponding quantities are $\vec{r}$, $\{x,y,z\}$,  $\{\rho,\vfi,z\}$. Frame origins are separated by $\vec{r}{\;^0}$, hence the coordinates are related by
\be
\vec{r}\;^{'} = \vec{r}+\vec{r}^{\;0};\qquad 
x\;^{'} = x+x^{\;0},\quad y\;^{'} = y+y^{\;0}, \quad z\;^{'} = z+z^{\;0}\; .
\label{coord_shift}
\ee
We  also use an alternative writing $x^{\;0}\equiv x_1^{\;0},\quad y^{\;0}\equiv x_2^{\;0},\quad x\equiv x_1,\quad y\;^{'}\equiv x_2^{'}$, etc.

The surfaces of the inner and outer cylinders are $\rho^{\;'}=a$ and $\rho=b$, respectively, with $d=b-a$. They carry arbitrary voltage distributions given by the conditions 
\be
\Phi\biggl|_{\rho^{\;'}=a}=G(\vfi^{\;'},z^{\;'}),\qquad \Phi\biggl|_{\rho=b}\;\;=\Vm+H(\vfi,z)\; ,
\label{bca}
\ee
where $\Phi$ is the electrostatic potential, and $\Vm$ is the uniform potential difference: all the voltages in the problem are counted from the uniform voltage of the inner cylinder taken as zero. The non--uniform potentials (patch patterns) are described by arbitrary smooth enough functions $G(\vfi^{\;'},z^{\;'})$ and $H(\vfi,z)$, whose local nature is emphasized by requiring
\be
||G||^2=\int\limits_{0}^{2\pi}\int\limits_{-\infty}^{\infty}d\vfi^{\;'}dz^{\;'}\,|G(\vfi^{\;'},z^{\;'})|^2<\infty,\qquad
||H||^2=\int\limits_{0}^{2\pi}\int\limits_{-\infty}^{\infty}d\vfi dz\,|H(\vfi,z)|^2<\infty\; .
\label{L2}
\ee

For any function $u(\vfi,z)$ satisfying the square integrability condition we denote its Fourier coefficient $u_n(k)$:
\bea
u(\vfi,z)=\LL\,u_n(k)e^{i(kz+n\vfi)},\quad u_n(k)=\FF\,u(\vfi,z)e^{-i(kz+n\vfi)}\; ; 
\label{Four}
\eea
in particular, Fourier coefficients of the patch voltages $G(\vfi^{\;'},z^{\;'})$ and $H(\vfi,z)$ are $G_n(k)$ and $H_n(k)$, respectively.
Since these functions are real, their Fourier coefficients satisfy
\be
G_n(k)=G^*_{-n}(-k),\qquad H_n(k)=H^*_{-n}(-k)\; ;
\label{symmGnHn}
\ee
here and elsewhere the star denotes complex conjugation. For any two squarely integrable functions $u(\vfi,z)$ and $v(\vfi,z)$ the useful Parceval identity holds:
\be
(u,\,v)\equiv\int\limits_{0}^{2\pi}\int\limits_{-\infty}^{\infty}d\vfi dz\,u(\vfi,z)v^*(\vfi,z)= 
\LLnof u_n(k)v^*_n(k)\; .
\label{Parc}
\ee
In the case $u=v$ the identity (\ref{Parc}) shows that the squared norm, $||u||^2$, of a function $u$ is equal to the squared norm of its Fourier coefficient $u_n(k)$.

As shown in CPEI, section III, electrostatic energy in the gap between the two cylinders, as a function of their mutual shift $\vec r^{\;0}$ consists of three parts,
\be\label{energy}
W(\vec r^{\;0})=W^u(\vec r^{\;0})+W^{int}(\vec r^{\;0})+W^{p}(\vec r^{\;0})\; ,
\ee
where the first one is due to the uniform potential difference, the second results from the interaction between the uniform and patch voltages, and the third one is the energy of the patch interaction. Each of these contributions was found in CPEI in the form of an expansion in the small transverse shift $\rho_0$ to quadratic order in the small $\rho_0/d$ ratio, with the coefficients depending generally on the axial shift $z_0$: 
\bea
W^{\cal A}(\vec r^{\;0})=W_0^{\cal A}(z^0)+W_\mu^{\cal A}(z^0)\,(x^0_\mu/d)+
\;W_{\mu\nu}^{\cal A}(z^0)\,(x^0_\mu/d)\,(x^0_\nu/d)+O\left[(\rho_0/d)^3\right]\; ;\nonumber\\
{\cal A}=u,\;int,\;p\;. \qquad\qquad\qquad\qquad\qquad\qquad\qquad\qquad\quad
\label{enexp}
\eea
The explicit coefficients for each kind of energy follow immediately.

The uniform potential energy is, naturally, proportional to the length of the capacitor, same as its expansion coefficients which, for $|z|\leq L$, are
\be\label{Wures}
W^u_0(L)=2\pi L \epsilon_0 \frac{a}{d}\left(\Vm\right)^2,\qquad
W^u_\mu(L)=0,\qquad
W^u_{\mu\nu}(L)=\pi L \epsilon_0 \frac{a}{d}\left(\Vm\right)^2\delta_{\mu\nu}\;.
\ee
Here and everywhere else we adopt the summation rule over repeated Greek indeces $\mu,\;\nu$, etc.: the summation over them runs from 1 to 2, corresponding to transverse coordinates in the cylinder cross--section. The first order term vanishes as it should be due to symmetry (otherwise there would be a non--zero transverse force in a perfectly symmetric configuration of coaxial conductors under uniform potentials).

The expansion coefficients for the interaction energy are:
\bea
W^{int}_0=-2\pi\epsilon_0\frac{a}{d}\Vm\left[
\left(G_0(0)-H_0(0)\right)
\right]\;,\nonumber\qquad\qquad\qquad\qquad\qquad\qquad\qquad\qquad\quad\,\\
\label{Wintres}
W^{int}_\mu=+4\pi\epsilon_0\frac{a}{d}\Vm\Re\left[c_\mu^+\left(G_{1}(0)-H_{1}(0)\right)
\right]\;,\qquad\qquad\qquad\qquad\qquad\qquad\qquad\;\;\;\;\\
W^{int}_{\mu\nu}=-4\pi\epsilon_0\frac{a}{d}\Vm\left\{\Re\left[c_\mu^+ c_\nu^+\left(G_{2}(0)-H_{2}(0)\right)\right]+(\delta_{\mu\nu}/4)\left(G_0(0)-H_0(0)\right)
\right\}\;,\quad\;\,\nonumber
\eea 
where $c_1^+=0.5, \;c_2^+=0.5i$, $\Re(\cdot)$ denotes the real part of $(\cdot)$; recall also that $G_n(k)$ and $H_n(k)$ are the Fourier coefficients of the patch voltages (\ref{bca}). Same as the uniform part of energy, the interaction one does not depend on the axial shift, because the longitudinal shifting of an electorde with the uniform potential does not change the actual charge configuration.

Finally, the patch energy coefficients are found to be:
\bea
\label{Wpres}
W^{p}_0=\frac{\epsilon_0 a}{2d}\int_{-\infty}^{\infty}dk\sum_{n=-\infty}^{\infty}|G_n(k)e^{\imath k z^0}-H_n(k)|^2\;;\nonumber\qquad\qquad\qquad\qquad\qquad\qquad\qquad\qquad\;\;\;\;\,\\
W^{p}_\mu=-\frac{\epsilon_0 a}{d}\int_{-\infty}^{\infty}dk\sum_{n=-\infty}^{\infty}\,\Re\left[c_\mu^+\left(G_n^{*}(k)e^{-\imath k z^0}-H_n^{*}(k)\right)\left(G_{n+1}(k)e^{\imath k z^0}-H_{n+1}(k)\right)\right]\;;\quad\;\,\\
W^{p}_{\mu\nu}=\frac{\epsilon_0 a}{2d}\int_{-\infty}^{\infty}dk\sum_{n=-\infty}^{\infty}\,\left\{
\delta_{\mu\nu}/2|G_n(k)e^{\imath k z^0}-H_n(k)|^2+\right.\nonumber\qquad\qquad\qquad\qquad\qquad\qquad\qquad\\
\left.2\Re\left[c_\mu^+c_\nu^+\left(G_n^{*}(k)e^{-\imath k z^0}-H_n^{*}(k)\right)\left(G_{n+2}(k)e^{\imath k z^0}-H_{n+2}(k)\right)
\right]
\right\}\nonumber\;.\;\;\;
\eea
These quantities do depend on the axial shift, $z^0$. Combining expressions (\ref{energy}) and (\ref{enexp}) we can also write for the total energy:
\be
W(\vec r^{\;0})=W_0(z^0)+W_\mu(z^0)\,(x^0_\mu/d)+\;W_{\mu\nu}(z^0)\,(x^0_\mu/d)\,(x^0_\nu/d)+O\left[(\rho_0/d)^3\right]\; ;
\label{totenexp}
\ee
\be
W_\xi(z^0)=W_\xi^u+W_\xi^{int}+W_\xi^p(z^0),\qquad \xi=0,\;\mu,\;\mu\nu\; .
\label{totencoef}
\ee
Note that all parts of energy are given to l. o. in $d/a$. 

\section{Patch Effect Forces\label{s3}}

As explained in the introduction, the force is found by the formulas (\ref{forcegen}) and (\ref{totenexp}):
\bea
F_\mu=-\frac{\partial W(\vec{r}{\;^0)}}{\partial x^0_\mu}=
-\left[
W_\mu(z^0)+2W_{\mu\nu}(z^0)\,(x^0_\nu/d)+O\left((\rho_0/d)^2\right)
\right]=\qquad\qquad\qquad\qquad\nonumber\\
-\left[
W_\mu(0)+W^{\;'}_\mu(0)\,z^0+2W_{\mu\nu}(0)\,(x^0_\nu/d)+O\left((r_0/d)^2\right)
\right],\quad \mu=1,2\; ;\qquad\quad\;\;\label{TrForceGen}\\
 r_0=|\vec{r}{\;^0}|\equiv\sqrt{\left(x^0\right)^2+\left(y^0\right)^2+\left(z^0\right)^2}\; ,\qquad\qquad\quad\nonumber
\eea
for the transverse force components, and
\bea
F_z=-\frac{\partial W(\vec{r}{\;^0)}}{\partial z^0}=
-\left[W^{'}_0(z^0)+
W^{\;'}_\mu(z^0)\,(x^0_\mu/d)+O\left((\rho_0/d)^2\right)
\right]=\nonumber\\
\qquad\qquad\qquad\qquad
-\left[
W_0^{\;'}(0)+W^{\;''}_0(0)\,z^0+W_{\mu}^{\;'}(0)\,(x^0_\mu/d)+O\left((r_0/d)^2\right)
\right]\;,\label{zForceGen}
\eea
for the axial force (here and everywhere else primes denote the derivatives in $z^0$). Same as energy, the force consists, of course of three parts, which we study below one by one.

\subsection{The Force due to the Uniform Potential Difference}\label{s3.1}

Using the formula (\ref{TrForceGen}) for the transverse force and expansion (\ref{enexp}) with the coefficients (\ref{Wures}) we obtain
(as usual, $F_1^u=F_x^u,\;F_2^u=F_y^u$):
\be
F_x^u=-2\pi L \epsilon_0 \frac{a}{d\,^2}\left(\Vm\right)^2 \left(x^0/d\right),\qquad F_y^u=-2\pi L \epsilon_0 \frac{a}{d\,^2}\left(\Vm\right)^2 \left(y^0/d\right),\qquad F_z^u=0\; .
\label{Funif}
\ee
Zero axial force, obvious by symmetry, is formally due to the independence of the energy  on the axial shift, $z^0$. Recall that $2L$ is the cylinder length, so the force per unit length of the cylinders that is finite. 

\subsection{The Patch and Uniform Potential Interaction Force}\label{s3.2}

To get it, we combine formulas (\ref{TrForceGen}), (\ref{enexp}) and (\ref{Wintres}); after some simplifying transformations the result becomes [$\Im(\cdot)$ is the imaginary part of $(\cdot)$]:
\bea
F^{int}_x=-2\pi\epsilon_0\frac{a}{d\,^2}\Vm\left\{
\Re\left[G_1(0)-H_1(0)\right]-\right.\nonumber\qquad\qquad\qquad\qquad\qquad\qquad\qquad\qquad\qquad\qquad\\
\left.
\Re\left[\left(G_2(0)-H_2(0)\right)+\left(G_0(0)-H_0(0)\right)\right]\left(x^0/d\right)+\Im\left[G_2(0)-H_2(0)\right]\left(y^0/d\right)
\right\}\;;\quad\qquad\qquad\label{Fint}\\
F^{int}_y=-2\pi\epsilon_0\frac{a}{d\,^2}\Vm\left\{
-\Im\left[G_1(0)-H_1(0)\right]+\Im\left[G_2(0)-H_2(0)\right]\left(x^0/d\right)+\right.\nonumber\qquad\qquad\qquad\quad\\
\left.
\Re\left[\left(G_2(0)-H_2(0)\right)-\left(G_0(0)-H_0(0)\right)\right]\left(y^0/d\right)
\right\}\;;\quad\qquad\qquad\qquad\qquad\qquad\nonumber\\
F^{int}_z=0\;.\qquad\nonumber\;\;\qquad\qquad\qquad\qquad\qquad\qquad
\eea
The axial force vanishes again, by symmetry.

\subsection{Forces due to the Patch Interaction\label{s3.3}}

By the formula (\ref{TrForceGen}) and the energy expansion (\ref{enexp}) with the coefficients (\ref{Wpres}) we find:
\bea
F^p_x=\frac{\epsilon_0 a}{2d\,^2}\int_{-\infty}^{\infty}dk\sum_{n=-\infty}^{\infty}\left\{
\Re\left[\left(G_n^{*}(k)e^{-\imath k z^0}-H_n^{*}(k)\right)\left(G_{n+1}(k)e^{\imath k z^0}-H_{n+1}(k)\right)\right]-\right.\nonumber\qquad\qquad\\
\left.
\left[|G_n(k)e^{\imath k z^0}-H_n(k)|^2+\Re\left[\left(G_n^{*}(k)e^{-\imath k z^0}-H_n^{*}(k)\right)\left(G_{n+2}(k)e^{\imath k z^0}-H_{n+2}(k)\right)
\right]\right]\left(x^0/d\right)+\right.\nonumber\\
\left.
\Im\left[\left(G_n^{*}(k)e^{-\imath k z^0}-H_n^{*}(k)\right)\left(G_{n+2}(k)e^{\imath k z^0}-H_{n+2}(k)\right)
\right]\left(y^0/d\right)
\right\}\;;\quad\label{Fptr}\\
F^p_y=-\frac{\epsilon_0 a}{2d\,^2}\int_{-\infty}^{\infty}dk\sum_{n=-\infty}^{\infty}\left\{
\Im\left[\left(G_n^{*}(k)e^{-\imath k z^0}-H_n^{*}(k)\right)\left(G_{n+1}(k)e^{\imath k z^0}-H_{n+1}(k)\right)\right]-\right.\nonumber\qquad\;\;\;\,\\
\left.
\Im\left[\left(G_n^{*}(k)e^{-\imath k z^0}-H_n^{*}(k)\right)\left(G_{n+2}(k)e^{\imath k z^0}-H_{n+2}(k)\right)
\right]\left(x^0/d\right)+\right.\nonumber\\
\left.
\left[|G_n(k)e^{\imath k z^0}-H_n(k)|^2-\Re\left[\left(G_n^{*}(k)e^{-\imath k z^0}-H_n^{*}(k)\right)\left(G_{n+2}(k)e^{\imath k z^0}-H_{n+2}(k)\right)
\right]\right]\left(y^0/d\right)
\right\}\;.
\nonumber
\eea
To get the axial force, we use formulas (\ref{enexp}), (\ref{Wpres}), and (\ref{zForceGen}):
\bea
F_z^p=-\frac{\epsilon_0 a}{d}\LLnof \label{Fpax}
\left\{
\Im\left[kG_n(k)H_n^*(k)e^{\imath k z^0}\right]+\right.\qquad\qquad\;\;\;\;\,\\
 \left.
\frac{\Im}{2}\left[G_n^*(k)H_{n+1}(k)e^{-\imath k z^0}-
H_n^*(k)G_{n+1}(k)e^{\imath k z^0}\right]k\left(x^0/d\right)\right.+\nonumber\\
\left.\frac{\Re}{2}\left[G_n^*(k)H_{n+1}(k)e^{-\imath k z^0}-
H_n^*(k)G_{n+1}(k)e^{\imath k z^0}\right]k\left(y^0/d\right)
\right\}\nonumber\;.
\eea

Formulas (\ref{Fptr}) and (\ref{Fpax}) provide the force due to patches to linear order in a small transverse shift for an arbitrary axial displacement. In many cases, such as the STEP set--up described in sec.~\ref{s6}, the axial shift  is also small, and PE forces to linear order in all the shifts are only needed. We obtain these expressions by replacing $\exp{(\pm\imath k z^0)}$ with the two terms of its Maclaurin expansion, assuming that all the arising integrals in $k$ converge:
\bea
F^p_x=\frac{\epsilon_0 a}{2d\,^2}\int_{-\infty}^{\infty}dk\sum_{n=-\infty}^{\infty}\left\{
\Re\left[\left(G_n^{*}(k)-H_n^{*}(k)\right)\left(G_{n+1}(k)-H_{n+1}(k)\right)\right]-\right.\nonumber\qquad\qquad\\
\left.
\left[|G_n(k)-H_n(k)|^2+\Re\left[\left(G_n^{*}(k)-H_n^{*}(k)\right)\left(G_{n+2}(k)-H_{n+2}(k)\right)
\right]\right]\left(x^0/d\right)+\right.\nonumber\\
\left.
\Im\left[\left(G_n^{*}(k)-H_n^{*}(k)\right)\left(G_{n+2}(k)-H_{n+2}(k)\right)
\right]\left(y^0/d\right)+
\right.\nonumber\\\left.
\Im(G_n(k)H_{n+1}^{*}(k)-H_n(k)G_{n+1}^{*}(k))\left(kz^0\right)
\right\}\;;\qquad\nonumber\\
F^p_y=-\frac{\epsilon_0 a}{2d\,^2}\int_{-\infty}^{\infty}dk\sum_{n=-\infty}^{\infty}\left\{
\Im\left[\left(G_n^{*}(k)-H_n^{*}(k)\right)\left(G_{n+1}(k)-H_{n+1}(k)\right)\right]-
\right.\label{Fpz0}\qquad\;\;\;\,\\
\left.
\Im\left[\left(G_n^{*}(k)-H_n^{*}(k)\right)\left(G_{n+2}(k)-H_{n+2}(k)\right)
\right]\left(x^0/d\right)+\right.\nonumber\\
\left.
\left[|G_n(k)-H_n(k)|^2-\Re\left[\left(G_n^{*}(k)-H_n^{*}(k)\right)\left(G_{n+2}(k)-H_{n+2}(k)\right)
\right]\right]\left(y^0/d\right)+\right.\nonumber\\
\left.
\Re(G_n(k)H_{n+1}^{*}(k)-H_n(k)G_{n+1}^{*}(k))\left(kz^0\right)
\right\}\;;\qquad\nonumber\\
F_z^p=-\frac{\epsilon_0 a}{d}\LLnof 
\left\{
\Im\left[kG_n(k)H_n^*(k)\right]+\right.\qquad\qquad\qquad\qquad\qquad\qquad\qquad\nonumber\\
 \left.
\frac{\Im}{2}\left[G_n^*(k)H_{n+1}(k)-
H_n^*(k)G_{n+1}(k)\right]k\left(x^0/d\right)\right.+\nonumber\\
\left.\frac{\Re}{2}\left[G_n^*(k)H_{n+1}(k)-
H_n^*(k)G_{n+1}(k)\right]k\left(y^0/d\right)+\Re(G_n(k)H_n^{*}(k))k\left(kz^0\right)
\right\}\nonumber\;.
\eea
Unlike the transverse shifts $x^0$ and $y^0$, the axial shift enters here not in the ratio to the gap, $d$, but in the product with $k$, which is the inverse characteristic length in the axial direction. 

Note that all the forces are derived as acting on the inner cylinder, the forces on the outer one have the opposite sign.

\subsection{General Properties of Electrostatic Patch Effect Forces\label{s3.4}}

The above results allow for some general conclusions regarding the patch interaction. 
\vskip1mm

1. To lowest order, the axial patch force is inversely proportional to the gap width, the transverse force components go as its inverse square.
\vskip1mm

2. Forces and shifts are directionally coupled: a transverse shift causes generally some axial force, and vice versa, a transverse force appears due to an axial shift.
\vskip1mm

3. The axial force vanishes when the patches are present on one of the cylinders only, i.e., when either $G_n(k)=0$, or $H_n(k)=0$. 
\vskip1mm

4. The transverse force does not vanish when the patches are on one of the cylinders only. Moreover, expressions (\ref{Fptr}) for its components can be given in terms of the non-zero patch potential (and not its Fourier coefficient!) using the Parceval identity (\ref{Parc}). For example, in the case when there are no patches on the inner cylinder $[\Va(\vfi^{\,'},z^{\,'})\equiv0, G_n(k)\equiv0]$ the transverse components are:
\bea
F^p_x=\frac{\epsilon_0 a}{2d\,^2}\int_{-\infty}^{\infty}dz\int_{0}^{2\pi}d\vfi\,\Vb^2(\vfi,z)
\left[
\cos\vfi-
\left(1+\cos2\vfi\right)\left(x^0/d\right)
-\sin2\vfi\left(y^0/d\right)
\right]\;;\nonumber\\
F^p_y=\frac{\epsilon_0 a}{2d\,^2}\int_{-\infty}^{\infty}dz\int_{0}^{2\pi}d\vfi\,\Vb^2(\vfi,z)
\left[
\sin\vfi-
\left(1-\cos2\vfi\right)\left(y^0/d\right)
-\sin2\vfi\left(x^0/d\right)
\right]\;.\nonumber\\
\nonumber
\eea
The transverse interaction force also does not vanish in such a case, by the formulas (\ref{Fint}).
\vskip1mm

5. Uniformly charged cylinders give rise to a mutual {\it restoring (elastic)} force, so that the coaxial condensor is at a neutrally stable equilibrium.
\vskip1mm

6. The interaction between patch and uniform potentials involves only the first and second polar angle harmonics of the patch distribution if the force is taken, as above, to linear order in the transverse shift.
\vskip1mm

Some of the above conclusions might be rather obviuos or intuitively clear, however, all of them are now accurately established by our analysis.

\section{The Patch Model\label{s4}}

To understand better patch interaction, it is natural to examine the case when just a couple of patches are present, and the PE forces are described by as simple expressions as possible. One thus needs some convenient {\it model} of a patch as a localized deviation from the uniform potential described by some particular functions with just few parameters involved. One of them should control the patch potential, two more have to govern the spot width in the axial and azimuthal directions, and two more parameters specify the patch position.

Such a patch model is desired for another reason as well. In the experiments there is usually no way to directly measure the patch distribution at the electrode surfaces. Instead, one should infer it from some other signals, like the patch forces. However, our force formulas are not fit for immediate modeling, since the unknowns in them are the Fourier coefficients $G_n(k)$ and $H_n(k)$, with no means to estimate these functions unless  properly parameterized. The existing experience of such parameterizations, along with the common sense, demonstrate clearly that only the models based on the underlying physics, rather than {\it ad hoc} ones, turn out efficient and work successfully.

So, the goal of an effective patch model is to find such functions that: a) both Fourier coefficients $G_n(k)$ and $H_n(k)$ are found in a closed form, and b) all the series and integrals in the formulas (\ref{Fptr}) and (\ref{Fpax}) for the forces are computed analytically in a closed form. From this standpoint, separation of variables is the simplest representation:
\be
\cV(\vfi-\vfi_*, z-z_*)=V_*\,f(z-z_*)\,u(\vfi-\vfi_*)\; .
\label{Vpatchgen}
\ee
Here the normalizing constant $V_*$ has the dimension of a potential, and the dimensionless functions $f(z)$ and $u(\vfi)$ are chosen so that
$|f(z)|\leq1,\; |u(\vfi)|\leq1; \quad f(0)=1,\quad u(0)=1$. The center of the patch is at $\vfi=\vfi_*,\;z=z_*$, where the potential achieves  the maximum magnitude $V_*$ (positive or negative). The Fourier coefficient of the function (\ref{Vpatchgen}) is:
\bea
\cV_n(k)=V_*\,\tf(k)e^{-ikz_*}\,u_n e^{-in\vfi_*}=V_*\,\tf(k)\,u_n e^{-i(kz_*+n\vfi_*)}\; ;\qquad
\label{VpatchgenFour}\\
\tf(k)=\frac{1}{\sqrt{2\pi}}\,\int_{-\infty}^\infty dz\, f(z)e^{-ikz},\qquad 
u_n= \frac{1}{\sqrt{2\pi}}\,\int_0^{2\pi} d\vfi\, u(\vfi)e^{-in\vfi}\; .\nonumber
\label{fkun}
\eea

Successful implementation of the above requirements a) and b) is in the choice of functions $f(z)$ and $u(\vfi)$. The first of them is natural to choose as the Gaussian exponent in $z$,
\be
f(z)=\exp{\left[-\left(\frac{z}{\sqrt{2}\,\Delta z}\right)^2\right]},\qquad
\tf(k)=\Delta z\exp{\left[-\left(\frac{k\Delta z}{\sqrt{2}}\right)^2\right]}\; ,
\label{fonk}
\ee
with the Fourier coefficient a Gaussian exponent, too, and the parameter $\Delta z$ giving the axial half--width of the spot. Since the product of two Gaussians entering the force expressions (\ref{Fptr}) and (\ref{Fpax}) is again a Gaussian, the integrals in $k$ there combinations of the elementary functions, as desired. A good choice of the second function, $u(\vfi)$, which is $2\pi$--periodic, turns out much more difficult. Nevertheless, eventually one can come up with the following:
\be
u(\vfi)=u(\vfi,\lambda)=\frac{\left(1-\lambda\right)^2}{2}\,\frac{1+\cos\vfi}{1-2\lambda\cos\vfi+\lambda^2}\;,
\label{uonfi}
\ee
where $-1\leq\lambda<1$ is controlling the width of the peak at $\vfi=0$ (see fig. \ref{fig2} below). Indeed, when $\lambda=-1$, $u(\vfi,-1)\equiv 1$, for $\lambda=0$, $u(\vfi,0)=0.5(1+\cos\vfi)$, and finally, when $\lambda\to1-0$, the function demonstrates a `bounded delta--like' behavior by going to zero everywhere except $\vfi=0$, where the limit is unity (zero width peak). We introduce the azimuthal patch half--width, $\Delta\vfi$, in an accurate way as the abscissa at the point where $u$ coincides with its mean value over the whole interval, $u(\Delta\vfi)=u_{av}$, which gives
\be
\cos\Delta\vfi={\lambda},\qquad\qquad\Delta\vfi=\arccos{\lambda}\; .
\label{Deltafi}
\ee
In a complete agreement with the above, when $\lambda\to 1-0$, the width shrinks according as 
$\Delta\vfi\approx\sqrt{2\left(1-\lambda\right)}\to 0$; in the opposite case $\lambda=-1$ one naturally has $2\Delta\vfi=2\pi$.

What makes the choice (\ref{uonfi}) really invaluable for calculations is its Fourier coefficients gotten by simply expanding the function in powers of $\lambda\exp{(-i\vfi)}$:
\be
u_n=u_n(\lambda)=\sqrt{2\pi}\,\frac{1-\lambda^2}{4\lambda}\,\lambda^{\left|n\right|},\quad n\neq 0;
\qquad\qquad u_0=u_0(\lambda)=\sqrt{2\pi}\,\frac{1-\lambda}{2}\; ,
\label{uonn}
\ee
which are essentially just exponents of $|n|$, as in the geometric progression. Apparently, expressions (\ref{uonfi}) and (\ref{uonn}) satisfy our requirements a) and b), perhaps even in the simplest possible way. The profiles of $u(\vfi)$ are plotted in fig. \ref{fig2} for various width values.

With $u(\vfi)$ and $f(z)$ defined by formulas (\ref{fonk}) and (\ref{uonfi}), our patch model (\ref{Vpatchgen}) becomes:
\bea\label{Vpatch}
\cV(\vfi-\vfi_*, z-z_*)=\cV(\vfi-\vfi_*, z-z_*;\,\Delta z,\Delta\vfi)\equiv V_*\, v(\vfi-\vfi_*, z-z_*;\,\Delta z,\Delta\vfi)=\qquad\\
\nonumber
V_*\,\frac{\left(1-\lambda_*\right)^2}{2}\frac{1+\cos{\left(\vfi-\vfi_*\right)}}{1-2\lambda_*\cos{\left(\vfi-\vfi_*\right)}+\lambda_*^2}\exp{\left[-\left(\frac{z-z_*}{\sqrt{2}\,\Delta z_*}\right)^2\right]}\; .\qquad\qquad\qquad\qquad\qquad\quad
\eea
The corresponding Fourier coefficients are found by the expressions (\ref{VpatchgenFour}), (\ref{fonk}), and (\ref{uonn}) as
\bea
\cV_n(k)=\sqrt{2\pi}\,\Delta z_*\, V_*
\frac{1-\lambda_*^2}{4\lambda_*}\,\lambda_*^{\left|n\right|}\,\exp{\left[-\left(\frac{k\Delta z_*}{\sqrt{2}}\right)^2\right]}e^{-i\left(n\vfi_*+kz_*\right)}\;,\;\; n=\pm1,\pm 2\ldots\,,\qquad\qquad\nonumber\\
\cV_0(k)=\sqrt{2\pi}\,\Delta z_*\, V_*\,\frac{1-\lambda_*}{2}\,
\exp{\left[-\left(\frac{k\Delta z_*}{\sqrt{2}}\right)^2\right]}e^{-ikz_*}\; .\qquad\qquad\qquad\qquad\qquad\qquad\qquad\quad
\label{VpatchFour}
\eea
Certainly, these functions are smooth enough to satisfy conditions (\ref{L2}), in fact, the function (\ref{Vpatch}) and all its derivatives in $\vfi$ and $z$ are squarely integrable, or else $\cV$ belongs to the Sobolev space $H_p$ for any $p\geq0$. The picture of equipotentials of the patch (\ref{Vpatch}) normalized by the maximum voltage is in fig. \ref{fig3}.

\section{Single Patch at Each of the Electrodes\label{s5}: a Picture of Patch Interaction}

We consider now two patches of the form (\ref{Vpatch}), one at the inner, the other at the outer boundary. Patch voltages in the boundary conditions (\ref{bca}) become ($i=1,\;2$):
\bea
\label{Onebcp}
G(\vfi^{'},z^{'})=\cV (\vfi^{'}-\vfi_1, z^{'}-z_1),\qquad 
H(\vfi,z)=\cV (\vfi-\vfi_2, z-z_2)\; ;\qquad\qquad\\
\cV (\vfi-\vfi_i, z-z_i)=V_i\,
\frac{\left(1-\lambda_i\right)^2}{2}
\frac{1+\cos\left(\vfi-\vfi_i\right)}{1-2\lambda_i\cos\left(\vfi-\vfi_i\right)+\lambda_i^2}\exp{\left[-\left(\frac{z-z_i}{\sqrt{2}\,\Delta z_i}\right)^2\right]}\; ,\nonumber
\eea
where, according to the relation (\ref{Deltafi}) between $\lambda$ and the angular width, $\Delta \vfi$, \break $0<\Delta \vfi_i\le \pi,\quad 0\le\Delta z_i< \infty,\quad -\pi<\vfi_i\le \pi,\quad -\infty< z_i < \infty,\quad i=1,2$. The forces corresponding to these distributions are derived in the appendix. We study here the interaction of two identical patches, $\Delta z_1=\Delta z_2=\Delta z$, $\Delta \vfi_1=\Delta \vfi_2=\Delta \vfi$, and $V_1=\pm V_2=V_0$. 

\subsection{Transverse Force\label{s5.1}}

\subsubsection{Transverse force due to patch and uniform potential interaction\label{5.1.1}}

By formula (\ref{OneFint}) of the appendix, the transverse force due to the interaction between the uniform potential and patches reduces to the following expressions:
\bea
\label{OneFintres}
\frac{F_x^{int}}{F_0}=
-\pi\sqrt{\frac{\pi}{2}}\frac{\Delta z}{a}\,\sin^2\Delta\vfi\Biggl\{\left(\cos{\vfi_1}\mp\cos{\vfi_2}\right)-\qquad\qquad\qquad\qquad\nonumber\\
\frac{x^0}{d}\left[\left(\cos{2\vfi_1}\mp\cos{2\vfi_2}\right)\cos\Delta\vfi+\frac{2(1\mp 1)}{1+\cos\Delta\vfi}
\right]-
\frac{y^0}{d}\left[\left(\sin{2\vfi_1}\mp\sin{2\vfi_2}\right)\cos\Delta\vfi\right]
\Biggr\}\;;\nonumber\\
\frac{F_y^{int}}{F_0}=
-\pi\sqrt{\frac{\pi}{2}}\frac{\Delta z}{a}\,\sin^2\Delta\vfi\Biggl\{\left(\sin{\vfi_1}\mp\sin{\vfi_2}\right)-\qquad\qquad\qquad\qquad\\
\frac{x^0}{d}\left[\left(\sin{2\vfi_1}\mp\sin{2\vfi_2}\right)\cos\Delta\vfi
\right]-
\frac{y^0}{d}\left[\frac{2(1\mp 1)}{1+\cos\Delta\vfi}-\left(\cos{2\vfi_1}\mp\cos{2\vfi_2}\right)\cos\Delta\vfi\right]
\Biggr\}\; .\nonumber
\eea
Here $F_0$ is the characteristic force defined as
\be
\label{intscalef}
F_0={\epsilon_0 V_0\,\Vm}(a/d)^{\,2}\;,
\ee
and $\Vm$ is the uniform voltage difference (\ref{bca}). The minus or plus sign is taken for the patch voltages of the same or opposite sign, respectively; the signs of the charges induced  by patches on the cylinders are opposite in the first case, and same in the second.  As expected, the magnitude of the transverse force is inversely proportional to $(a/d)^2$. It is also proportional to the relative axial width, $\Delta z/a$, and  entirely independent of the axial positions $z_{1,2}$ of the patches. The dependence on the angular width is more complicated: the maximum force is at $\Delta\vfi=\pi/2$, as prompted by geometry; for a small width the force goes to zero as $(\Delta\vfi)^2$. When $\Delta\vfi=\pi$, i.e., the patches are the `belts' of voltage uniform in $\vfi$, the zeroth order force vanishes, and the total becomes proportional to the shift and directed along it:
\be
\vec F^{int}=(2\pi)^{3/2}(1\mp 1)F_0\,\left(\frac{\Delta z}{a}\right)\left(\frac{\vec\rho_0}{d}\right) \;. 
\label{belts}
\ee
This is zero when the patch voltages are equal: the forces from each of them have the same magnitude and opposite directions (see more on this below).

The main contribution, i.e., the force in the centered position is best characterized by its polar components, namely:
\bea
\label{trFintpolar}
\frac{F_\rho^{int}}{F_0}=
-\pi\sqrt{\frac{\pi}{2}}\frac{\Delta z}{a}\,\sin^2\Delta\vfi\left[\cos(\vfi-\vfi_1)\mp\cos(\vfi-\vfi_2)\right]
\;;\nonumber\\
\frac{F_\vfi^{int}}{F_0}=
\pi\sqrt{\frac{\pi}{2}}\frac{\Delta z}{a}\,\sin^2\Delta\vfi\left[\sin(\vfi-\vfi_1)\mp\sin(\vfi-\vfi_2)\right]
\;.\nonumber
\eea
The total force is a superposition of the two forces from each of the patches. They act along the radial direction to the corresponding patch center, and the total is a vector sum of these two radial vectors, see figs. \ref{fig4ab}. That is why the interaction force vanishes when the patches are one opposite the other ($\vfi_1=\vfi_2$): their contributions, aligned and of the opposite signs, exactly cancel each other. So here a patch behaves as an effective point charge, $q_{eff}$, in the uniform radial field $E^u=-\Vm/d$: the force due to it is just $q_{eff}E^u$. The effective charges are readily found by comparison with the above expressions of the force. 

The forces of the first order are directionally coupled to the shifts, meaning an $x$--force depends on the $y$--shift, and vice versa. The forces consist of a constant term and the second harmonics of the patch angular position. 

\subsubsection{Transverse force due to patch interaction\label{5.1.2}}

Its general expressions (\ref{OneFpxres}) and (\ref{OneFpyres}) simplify for the same size patches:
\bea
\label{OneFptrres}
\frac{F_x^p}{F_{tr}}=2(\pi)^{3/2}\frac{\Delta z}{a}\,\sin^2\left(\frac{\Delta\vfi}{2}\right)\Biggl\{
\left[
{\cal N}_1\mp 2{\cal{M}}_1\exp{\left[-\left(\frac{z_1-z_2}{2\Delta z}\right)^2\right]}
\right]\left(\cos\vfi_1+\cos\vfi_2\right)-\nonumber\qquad\\
\frac{x^0}{d}\left[
2{\cal N}_0+{\cal N}_2\left(\cos2\vfi_1+\cos2\vfi_2\right)\mp 2\Biggl({\cal M}_0+{\cal {M}}_2\cos{\left(\vfi_1+\vfi_2\right)}
\Biggr)\exp{\left[-\left(\frac{z_1-z_2}{2\Delta z}\right)^2\right]}
\right]-\nonumber\\
\frac{y^0}{d}\left[
{\cal N}_2\left(\sin2\vfi_1+\sin2\vfi_2\right)\mp 
2{\cal {M}}_2\sin{\left(\vfi_1+\vfi_2\right)}\exp{\left[-\left(\frac{z_1-z_2}{2\Delta z}\right)^2\right]}
\right]\mp \nonumber\\
\frac{z^0}{\Delta z}\left[
{\cal {M}}_1\left(\cos\vfi_1+\cos\vfi_2\right)\frac{z_1-z_2}{\Delta z} \exp{\left[-\left(\frac{z_1-z_2}{2\Delta z}\right)^2\right]}
\right]
\Biggr\}\;;\qquad\nonumber
\eea
\bea
\frac{F_y^p}{F_{tr}}=2(\pi)^{3/2}\frac{\Delta z}{a}\,\sin^2\left(\frac{\Delta\vfi}{2}\right)\Biggl\{
\left[
{\cal N}_1\mp 2{\cal{M}}_1\exp{\left[-\left(\frac{z_1-z_2}{2\Delta z}\right)^2\right]}
\right]\left(\sin\vfi_1+\sin\vfi_2\right)-\qquad\\
\frac{x^0}{d}\left[{\cal N}_2\left(\sin2\vfi_1+\sin2\vfi_2\right)\mp 2{\cal {M}}_2\sin{\left(\vfi_1+\vfi_2\right)}\exp{\left[-\left(\frac{z_1-z_2}{2\Delta z}\right)^2\right]}
\right]-\nonumber\\
\frac{y^0}{d}\left[2{\cal N}_0-{\cal N}_2\left(\cos2\vfi_1+\cos2\vfi_2\right)\mp 2\Biggl({\cal M}_0-{\cal {M}}_2\cos{\left(\vfi_1+\vfi_2\right)}\Biggr)\exp{\left[-\left(\frac{z_1-z_2}{2\Delta z}\right)^2\right]}
\right]\mp \nonumber\\
\frac{z^0}{\Delta z}\left[
{\cal {M}}_1\left(\sin\vfi_1+\sin\vfi_2\right)\frac{z_1-z_2}{\Delta z} \exp{\left[-\left(\frac{z_1-z_2}{2\Delta z}\right)^2\right]}
\right]
\Biggr\}\; .\nonumber\qquad
\eea
Here the characteristic transverse force, $F_{tr}$, is almost as in the previous case [formula (\ref{intscalef})],
\be
\label{pscalef}
F_{tr}={\epsilon_0 V_0^2\,}a^2/d^{\,2}\;,
\ee
with just a natural replacement of $\Vm$ with $V_0$. The coefficients involved are found by combining the formulas (\ref{Ms}), (\ref{Di}) and (\ref{Ns}) of the appendix with the representation (\ref{NewCoeff}):
\[
{\cal N}_0=\frac{3-\lambda}{8}\;;\qquad
{\cal N}_1=\frac{1+\lambda}{8}(2-\lambda)\;;\qquad
{\cal N}_2=\frac{1+\lambda}{16}(1+4\lambda-3\lambda^2)\;;
\]
\bea
\label{samecoeff}
{\cal M}_0=\frac{1-\lambda}{4}\left[1+\left(1+\lambda\right)^2\frac{\cos(\vfi_1-\vfi_2)-\lambda^2}{2D}\right]\;;\quad D=1-2\lambda^2\cos{\left(\vfi_1-\vfi_2\right)}+\lambda^4\;;\quad\\
{\cal M}_1=\frac{1-\lambda^2}{8}\left[1-\lambda\left(1+\lambda\right)
\frac{1+\lambda^2-2\cos(\vfi_1-\vfi_2)}{2D}\right]\nonumber\;;\qquad\qquad\qquad\qquad\qquad\qquad\quad\\
{\cal M}_2=\frac{1-\lambda^2}{8}\left\{\frac{1+\lambda}{2}+\right.
\qquad\qquad\qquad\qquad\qquad\qquad\qquad\qquad\qquad\qquad\nonumber\qquad\qquad\qquad\;\;\\
\left.
\lambda\left[2\cos{\left(\vfi_1-\vfi_2\right)}+
\lambda\left(1+\lambda\right)
\frac{\cos2(\vfi_1-\vfi_2)-\lambda^2\cos(\vfi_1-\vfi_2)}{D}\right]
\right\}\nonumber\;.\quad
\eea
As before, the signs $\mp$ correspond to the case of the same or opposite signs of the patch voltages.
The force (\ref{OneFptrres}) is again proportional to $(a/d)^2$ and $\Delta z/a$. The dependence on the angular width here is even more complicated than in the previous case; still, the force is $\propto(\Delta\vfi)^2$ in the narrow patch limit. For $\Delta\vfi=\pi$, an analog of the formula (\ref{belts}) holds:
\[
\vec F^{p}=2(\pi)^{3/2}F_{tr}\,
\left(\frac{\Delta z}{a}\right)\,\left\{1\mp \exp{\left[-\left(\frac{z_1-z_2}{2\Delta z}\right)^2\right]}\right\}\left(\frac{\vec\rho_0}{d}\right) \; . 
\label{beltsagain}
\]
This force, however, does not vanish for identical voltages, unless  the patches are one right against the other. 

The nature of the force due to the patch interaction is most clearly seen in the main term corresponding to coaxial cylinders with no axial shift:

\noindent a) \underline{$V_1=V_2=V_0$}
\[
\label{OneFptrcoax}
\frac{F_\bot^p}{F_{tr}}=
4(\pi)^{3/2}\,\frac{\Delta z}{a}\,\sin^2\left(\frac{\Delta\vfi}{2}\right)\Biggl|\,\cos{\frac{\vfi_1-\vfi_2}{2}}\Biggr|\left\{
{\cal N}_1-2{\cal M}_1\exp\left[-\left(\frac{z_1-z_2}{2\Delta z}\right)^2\right]
\right\}\;,
\]
\noindent b) \underline{$V_1=-V_2=V_0$}
\[
\frac{F_\bot^p}{F_{tr}}=
4(\pi)^{3/2}\,\frac{\Delta z}{a}\,\sin^2\left(\frac{\Delta\vfi}{2}\right)\Biggl|\,\cos{\frac{\vfi_1-\vfi_2}{2}}\Biggr|\left\{
{\cal N}_1+2{\cal M}_1\exp\left[-\left(\frac{z_1-z_2}{2\Delta z}\right)^2\right]
\right\}\;.\qquad
\]
The direction in the cylinder cross--section is along the bisectrix of the angle subtended by the patches, no matter what the voltage signs are:$\tan{\theta}^p=\tan[0.5(\vfi_1+\vfi_2)]$.
Thus for $\vfi_1=\vfi_2$ the total force is maximum because the contribution of each patch doubles, while for $\vfi_1=\vfi_2+\pi$ the total force is apparently zero. The force is a combination of two terms: one is proportional to ${\cal N}_1$ depending only on the angular patch width and non--negative, by the formulas (\ref{samecoeff}). The other is the exponent of $(z_1-z_2)^2$ with the coefficient ${\cal M}_1$ depending on $|\vfi_1-\vfi_2|$. The coefficient ${\cal M}_1$, along with ${\cal M}_0,$ and ${\cal M}_2$, are shown in fig. \ref{fig5} versus the angular distance. They have a sharp maximum at $\vfi_1=\vfi_2$, and drop quickly away from it; the maximum is the sharper, and the drop the faster, the smaller the width of the patch is. This is a strong manifestation of screening of patch charges: the patches interact strongest of all when their centers are right opposite each other; the interaction drops when one charge stops `seeing' the other due to the obstruction by the inner cylinder. 

In fig. \ref{fig6} $F_\bot^p$ is plotted as a function of the angular distance $|\vfi_1-\vfi_2|$ for both cases, a) and b). The two curves differ significantly for the moderate angular separations, and tend to zero when $\vfi_1-\vfi_2\to\pi$. The dependence of $F_\bot^p$ on the axial distance $|z_1-z_2|$  is shown in fig. \ref{fig7}.  The force does not vanish at infinity, but goes instead to the asymptotic value common for the two cases. 

As prompted by the similarity between the coefficients ${\cal M}_0$---${\cal M}_2$ and ${\cal N}_0$---${\cal N}_2$, being just some bounded functions of $\Delta \vfi$, the terms in the force (\ref{OneFptrres}) proportional to the transverse shifts have the structure, and thus the behavior, similar to that of the zeroth order expressions. The term with the axial shift is rather different, first of all because of an additional factor, $(d/\Delta z)$. Thus $z^0$ no longer compares to the gap, $d$: quite naturally, it is the ratio $(z^0/\Delta z)$ that stands as a small parameter at this part of the force. The other peculiarity is  the additional factor proportional to $z_1-z_2$. This part of the force vanishes in both limits, $(z_1-z_2)/\Delta z \to \infty$ and $(z_1-z_2)/\Delta z=0$, with the maximum magnitude at $|z_1-z_2|=\sqrt{2}\,\Delta z$ as seen in fig.\ref{fig8}. Finally, it is proportional to the first harmonics of the angular patch positions  and the coefficient ${\cal M}_1$. The coefficient, and thus the force, is maximum when $\vfi_1=\vfi_2$, then it decreases monotonically as $|\vfi_1-\vfi_2|$ increases, fig. \ref{fig5}.

\subsection{Axial Force\label{s5.2}}

As demonstrated in section \ref{s3}, the axial PE force is only due to the patch interaction. Its general expression (\ref{OneFpaxres}) reduces in our case to:
\bea
\label{sameOneFpaxres}
\frac{F_z^p}{F_{ax}}=\pm\,2\pi^{3/2}\,\sin^2{\left(\frac{\Delta\vfi}{2}\right)}\,\frac{z_1-z_2}{\Delta z}\exp\left[-\left(\frac{z_1-z_2}{2\Delta z}\right)^2\right]\Biggl\{{\cal M}_0-\qquad\qquad\qquad\qquad\qquad\\
\frac{x^0}{d} \,{\cal M}_1\left(\cos \vfi_1+\cos \vfi_2\right)-
\frac{y^0}{d} {\cal M}_1\left(\sin \vfi_1+\sin \vfi_2\right)-
\frac{z^0}{z_1-z_2}{\cal M}_0 \left[1-\left(\frac{z_1-z_2}{\sqrt{2}\Delta z}\right)^2\right]
\Biggr\}\; .\nonumber
\eea
The coefficients ${\cal M}_0,\,{\cal M}_1$ are found in the formulas (\ref{samecoeff}), and the characteristic force is inversely proportional to the relative gap, $d/a$, and not to its square, as before: 
\be
\label{z0pscalef}
F_{ax}=\epsilon_0 V_0^2 a/d\;. 
\ee
The sign of the force (\ref{sameOneFpaxres}) switches from plus to minus for patch voltages of the same or opposite signs, respectively. The most striking feature of the axial force is its overall dependence on the ratio $\left({z_1-z_2}\right)/{\Delta z}$.  As one expects intuitively, the axial force tends to zero in both limits $\Delta z\to0$ and $\Delta z\to\infty$, due to the Gaussian exponent of the above argument multiplied by this same argument. It has the maximum $\pi^{3/2}e^{-2}\,\sin^2{\left({\Delta\vfi}/{2}\right)}\,F_{ax}$ at $|z_1-z_2|=\sqrt{2}\,\Delta z$, same as the axial shift part of the transverse force from fig. \ref{fig8}.

The overall dependence of the axial force on $\Delta \vfi$ is just like that of the transverse force; particularly, for belt--like patches, $\Delta \vfi=\pi$, the force is: 
\[
\label{lastbelts}
\frac{F_z^p}{F_{ax}}=
\pm\,\pi^{3/2}\,\frac{z_1-z_2}{\Delta z}\,\exp\left[-\left(\frac{z_1-z_2}{2\Delta z}\right)^2\right]\Biggl\{ 
1-\frac{z^0}{z_1-z_2}\left[1-\left(\frac{z_1-z_2}{\sqrt{2}\Delta z}\right)^2\right] 
\Biggr\}\;.
\]
Here the main term vanishes when the two patches are in the same cross--section, $z_1=z_2$, but the `correction' proportional to $z^0$ does not:
${F_z^p}/{F_{ax}}=\mp\,{\pi^{3/2}}\,({z^0}/{\Delta z}),\;\;  \Delta \vfi=\pi,\quad z_1=z_2$. 

The first order force proportional to the axial shift has a factor $(z_1-z_2)^2$, instead of $(z_1-z_2)$, in front of the Gaussian. This contribution changes sign at $z_1-z_2=\sqrt{2}\,\Delta z$, and is maximum when the patches are in the same $z$ plane, and it as illustrated by fig. \ref{fig9}.

The axial force decreases with the angular distance between the patches. The zeroth order component of the axial force together with the first order term proportional to the axial shift depend on $|\vfi_1-\vfi_2|$ through the coefficient ${\cal M}_0$ only. The two other components due to the transverse shifts  are proportional to the harmonics of the patches angular positions and the coefficient ${\cal M}_1$; they both vanish in just one case, when $\vfi_1-\vfi_2=\pi$. Despite these minor differences, each of these terms follows basically the characteristic behavior of the coefficients ${\cal M}_0$ or ${\cal M}_1$ versus the distance $|\vfi_1-\vfi_2|$, as presented in fig. \ref{fig5}. 

\section{Estimates of Axial Patch Effect Force for STEP\label{s6}.\\ Concluding Remarks}

\subsection{ Basics of the STEP Experimental Set-up\label{s6.1} and Some Requirements}

STEP, a medium size scientific satellite ($<900\, Kg$), will be put into a drag--free earth orbit at the altitude of $\sim550\,Km$. It is to measure the relative free fall acceleration of pairs of test masses (TM) of different materials with an accuracy of $10^{-17}m/sec^2$, to determine the equivalence of the inertial and gravitational mass with an uncertainty 6 orders of magnitude smaller than the existing results \cite{Will}, or find violations of the Equivalence Principle (EP; in other words, Universal Free Fall) somewhere between 1 part in $10^{12}$ and 1 part in $10^{18}$.

STEP will fly four differential accelerometers (DAC), each with a pair of TMs shaped as coaxial cylindrical shells. The cross--section of the DAC is shown in fig. \ref{fig10}. An electromagnetic system of magnetic bearings and capacitors keeps TMs alligned and centred to within $<1 nm$. The transverse degrees of freedom are constrained, the two axial ones are left free, so the axial motion and the rotation about the TM axis are most important. So we discuss only the axial PE force; the axial torque will be examined in the final part of this paper.

The readout of the differential longitudinal displacement for each TM pair is provided by a Superconducting Quantum Interference Device (SQUID), sensitive to an acceleration of $3\times10^{-18}m/sec^2$ in an accumulation time of at least 20 orbits, i.e., $\sim2\,days$. Nominally the DACs are kept inertially fixed and the EP violation signal will be at the orbital frequency, $f_{orb}=1.74\times10^{-4}\,Hz$. However, changing the signal frequency from one measurement session to the other helps to discover and remove systematic readout errors. So modulation of the frequency is planned by rolling the satellite (and the DACs) about the normal to the orbital plane at some frequency $f_{roll}\geq 2f_{orb}$ (the DAC axes lie in the orbital plane). With this procedure, the science signal will be at the frequency $f_s=f_{roll}\pm f_{orb}$ bounded from below as
\be
f_{s}\geq 1.74\times10^{-4}\, Hz\; .
\label{frrang}
\ee

The STEP design rejects all perturbations able to mimic or mask the signal from the TM free fall at the level of the target accuracy. Thus the magnitude of any perturbing axial acceleration  at the signal frequency should be at least less than $10^{-17}m/sec^2$. In addition, axial motion of TMs is allowed only within certain range, as required by drag--free control and SQUID readout limits. The total axial shift must not exceed some limit, $z^0_{max}$, whose value may be adjusted while developing the instrument; currently it is $z^0_{max}= 1\,\mu m$.

The SQUID sensor will keep the TM position in the body by applying a restoring (spring) force to counteract any D.C. axial force, such as the one due to patches. Its control authority in terms of the range of the corresponding oscillation period is $300 - 1000\, s$. For a TM of $1\,Kg$, this converts into the spring constant, $k$, within the range $3.9\times10^{-5} - 4.4\times10^{-4}\,N/m$. Its maximum value $k_{max}=4.4\times10^{-4}\,N/m$ provides the maximum force that can be controlled according to $F_{max}=k_{max}z^0_{max}=4.4\times10^{-10}\,N$, or maximum acceleration
\be
\label{amax}
a_{max}=F_{max}/1\,Kg=4.4\times10^{-10}\,m/s^2\;.
\ee
The related criterion is: any distribution of patches is acceptable as soon as the axial D.C. acceleration due to them does not exceed $a_{max}$.

We consider a TM and its magnetic bearing as a reference case of our pair of cylinders (the gap between them is at least 3 times smaller - and the force equally larger - than the gap between the TM and electrodes, see fig. \ref{fig10}). First we assume each cylinder to carry just one patch of the same sizes and magnitudes; then we consider an arbitrary number of such patch  pairs. Using slightly different notations the formula (\ref{sameOneFpaxres}) for the axial force with $x^0=y^0=0$ becomes: 
\bea
\label{estFzp}
{F_z}=\pm{\pi^{3/2}}F_{ax}\sin^4{\left(\frac{\Delta\vfi}{2}\right)}\mu\exp\left[-\left(\frac{z_1-z_2}{2\Delta z}\right)^2\right]\Biggl\{\frac{z_1-z_2}{\Delta z}-
\frac{z^0}{\Delta z}\left[1-\left(\frac{z_1-z_2}{\sqrt{2}\Delta z}\right)^2\right]
\Biggr\}\;;\quad\\
\nonumber
\mu=\mu(\lambda, |\vfi_1-\vfi_2|)=1+\frac{\left(1+\lambda\right)^2}{2}\,\frac{\cos(\vfi_1-\vfi_2)-\lambda^2}{1-2\lambda^2\cos{\left(\vfi_1-\vfi_2\right)}+\lambda^4},
\quad \lambda=\cos\Delta\vfi\; ;\qquad
\eea
the characteristic force $F_{ax}$ is defined by the expression (\ref{z0pscalef}). To l. o. the force (\ref{estFzp}) is constant (unless the patches move on the surfaces, which has not been observed so far, or the cylinders rotate); the correction proportional to the axial shift, $z^0$, can produce harmonic oscillations near a stable equilibrium, or exponential runaway from an unstable one.   
We take the patch voltage $V_0=10\, mV$ for our estimates. This seems a plausible number for patches at low temperatures (see some relevant result for the GP-B experiment in~\cite{BT}). The TM parameters used are: TM radius $a=2.3\,cm$ (outer TM; the inner radius, and hence the force, is about 5 times smaller), TM height $2L=14\,cm$, the TM to magnetic bearing gap $d=0.3\,mm$ ($d/a\approx10^{-2}$). The TM masses vary from $0.3\,Kg$ to $2.4\, Kg$, so to make the rescaling easy, we give all the accelerations (or specific forces) {\it per 1 Kg of mass}.

\subsection{Constant Axial Acceleration due to Patches\label{s6.2}}

By the expression (\ref{estFzp}), the constant part of the axial patch effect force is
\be
\label{Fconst}
{F_z}=\pm\,2\pi^{3/2}F_{ax}\sin^4{\left(\frac{\Delta\vfi}{2}\right)}\mu\,
\tilde{z}\,\exp\left(-{\tilde{z}}^2\right),\quad
\tilde{z}\equiv(z_1-z_2)/2\Delta z\; ;
\ee
the ballpark number for the resulting acceleration comes from the characteristic force (\ref{z0pscalef}): 
\be
\label{aax}
a_{ax}=F_{ax}/1\,Kg=\epsilon_0{V}_0^2\, (a/d)/1\,Kg\approx 6.8\times10^{-14}\,m/s^2\; .
\ee
It is almost 4 orders of magnitude smaller than the one required by the formula (\ref{amax}). Next is an upper bound on the acceleration from the force (\ref{Fconst}) with two last factors set at their maximum. Since the maximum of $\mu$ with regards to $|\vfi_1-\vfi_2|$ is 
$\left[1+\sin^2{\left({\Delta\vfi}/{2}\right)}\right]/2\sin^2{\left({\Delta\vfi}/{2}\right)}$, and the last factor's maximum is $1/\sqrt{2e}$ when $\tilde{z}=1/\sqrt{2}$, the bound is
\be
\label{maxestazp}
|a_z|<\sqrt{\frac{2\pi^3}{e}}\,a_{ax}\sin^2\left(\frac{\Delta \vfi}{2}\right)
\approx3.2\times10^{-13}\,\sin^2\left(\frac{\Delta \vfi}{2}\right)\,m/s^2\;.
\ee
It is still more than 3 orders of magnitude smaller than the required value (\ref{amax}) for any $\Delta \vfi$, and drops as its square when it becomes small. 

We now consider the mean value of the force (\ref{Fconst}) over all angular $|\vfi_1-\vfi_2|$ and the axial, $\tilde{z}$ patch distances ($l_0\equiv L/2\Delta z$). The average in the angle is unity while that in the distance is $(l_0)^{-1}\int^{l_0}_0 \tilde z \exp(-\tilde z^2)d \tilde{z}=(l_0)^{-1}\left[1-\exp(-l_0)^2\right]$, so 
\bea
\label{meanestazp}
\bar{a}_z=
2{\pi^{3/2}}{a_{ax}}\frac{\Delta z}{L}\left[1-e^{-(L/2\Delta z)^2}\right]\sin^4\left(\frac{\Delta \vfi}{2}\right)
<
7.6\times10^{-13}\frac{\Delta z}{L}\sin^4\left(\frac{\Delta \vfi}{2}\right)\,m/s^2 \;.
\eea
Two new pleasant features here as compared to the estimate (\ref{maxestazp}) are: the fourth power instead of the square of sine, and a factor  $(\Delta z/L)$, which is less than unity in any case, and is expected to be essentially smaller. The upper bound (\ref{meanestazp}) satisfies the condition (\ref{amax}) with the margin of 3 orders of magnitude for any patch sizes.

However, this is hardly the case with some number, $N$, of patches pairs identical in sizes and magnitude, since $N$ may be quite large. Those pairs generate $N^2$ interactions each giving the force (\ref{Fconst}), with either sign. Assuming the signs of the patches random, single contributions do not cancel each other completely, but sum up to a factor $\sqrt{N^2}=N$ for $N\gg1$. Thus the total force is reasonably estimated as $N$ times the average of the force (\ref{Fconst}). The resulting acceleration satisfies the condition (\ref{amax}) if the following inequality is true:
\be
\label{cond2}
N\,\frac{\Delta z}{L}\sin^4\left(\frac{\Delta \vfi}{2}\right)<5.8\times10^{2} \; .
\ee
For $N$ up to 600 the condition is satisfied by the patches of any sizes. Limitations on the latter start with $N\sim1000$, they are not too restrictive until $N\sim10,000$: even for this number and the angular size $\Delta\vfi=60^\circ$, any axial size of the patch is still acceptable.

Moreover, the upper bound for $N$ is available through the patch sizes, if the patches do not overlap within their nominal widths $2\Delta z$,  $2\Delta\vfi$. The effective patch area is $(2\Delta z)\times (2a\Delta\vfi)$, the total surface area of the cylinder is $(2\pi a)(2L)$, so $N$ is bounded by the ratio
\be\label{Nmax}
N\leq\frac{4\pi aL}{4(a\Delta\vfi)\Delta z}=\frac{\pi }{\Delta\vfi}\,\frac{L}{\Delta z}\; .
\ee
Introducing this to the inequality (\ref{cond2}), we obtain a universal estimate in terms of $\Delta\vfi$ only:
\be
\label{cond3}
\frac{1}{\Delta\vfi}\,\sin^4\left(\frac{\Delta \vfi}{2}\right)<180 \; ,
\ee
which, of course, is always true. So, non-overlapping patches covering the surfaces completely satisfy the STEP limitation on the DC acceleration, if the averaged force is used and the assumption of random signs of the patch potentials is valid.

\subsection{Harmonic Oscillations and Exponential Runaway due to Patches\label{s6.3}}

The zero order force (\ref{estFzp}) vanishes at $z_1=z_2$. The first correction to it, $\delta F_{z}$, is:
\be
\label{Flin}
\delta F_{z}=\mp{\pi^{3/2}}\frac{F_{ax}}{\Delta z}\sin^4{\left(\frac{\Delta\vfi}{2}\right)}\mu\,
\left(1-2{\tilde{z}^2}\right)\,\exp\left(-{\tilde{z}}^2\right),\quad
\tilde{z}\equiv(z_1-z_2)/2\Delta z\; .
\ee
As explained in sec. \ref{s5.2}, its overall sign depends on the signs of the patches and the distance between them, because the factor in brackets is positive or negative depending on whether $\tilde{z}$ is smaller or larger than $1/\sqrt2.$ So we denote $\omega^2=\left|\delta F_{z}\right|/1\,Kg$, then the equation of TM motion in the axial direction near the equilibrium becomes:
\be
\label{zequatmot}
\ddot{z^0}=\mp\omega^2z^0\;,
\ee
and describes small oscillations for the minus sign (restoring force), and exponential runaway otherwise. In the first case, the oscillation frequency is $f=\omega/2\pi>0$, in the second the TM drifts away exponentially with the characteristic time $\tau=1/\omega$. 

The maximum of $\omega$, needed for an upper bound for the frequency $f$ or a lower bound for the time constant $\tau$, is clearly attained when $\tilde z=\vfi_1-\vfi_2=0$, so:
\be
\label{maxomega}
\omega\leq\sqrt{{\pi^{3/2}}\frac{a_{ax}}{\Delta z}\sin^2{\left(\frac{\Delta\vfi}{2}\right)}}=6.2\times 10^{-7}\frac{1}{\sqrt{\Delta z}}\sin\left(\frac{\Delta\vfi}{2}\right)\,rad/s\;,
\ee
with $\Delta z$ in meters. Remarkably, this estimate of $\omega$ decreases when $\Delta\vfi$ goes down, but increases with the decrease of $\Delta z$. 
The frequency of TM oscillations should be below the minimum science signal frequency (\ref{frrang}), 
\be
\label{frcrit}
f<1.74\times 10^{-4}\,Hz\;,
\ee
to avoid the signal corruption. Based on the estimate (\ref{maxomega}), this condition holds when
\be
\label{cond5}
\Delta z>3.2\times 10^{-7}\sin^2\left(\frac{\Delta\vfi}{2}\right)\,m\;.
\ee
Even if $\sin\left({\Delta\vfi}/2\right)=1$ (circular patches), the axial width should be only about two hundreds nanometers to satisfy this. 

In the runaway case the characteristic time must satisfy 
\be\label{taucrit}
\tau \gg T_{obs}=1.7\times10^{5}\,s\; ,
\ee 
to avoid saturation: if it is true, then any initial shift smaller than $z^0_{max}=1\,\mu m$ will not grow critically during a science session. This criterion and the estimate (\ref{maxomega}) lead to
\be
\label{cond6}
\Delta z\gg 1.1\times 10^{-2}\sin^2\left(\frac{\Delta\vfi}{2}\right)\,m\; ,
\ee
five orders of magnitude more restrictive than the bound (\ref{cond5}). However, a small enough angular size helps to meet the requirement (\ref{taucrit}). The maximum acceleration allowed in the runaway case when the condition (\ref{taucrit}) holds is implied by the estimate (\ref{maxomega}) with $\Delta z_{min}$ taken from the r.h.s of the estimate (\ref{cond6}) and the maximum shift $z^0_{max}=1\,\mu m$:
\be
\label{limitacc}
a^{max}=\omega^2z^0_{max}=3.8\times 10^{-19}\;\frac{\sin^2\left({\Delta\vfi}/{2}\right)}{\Delta z_{min}}\,m/s^2=
3.4\times10^{-17}\,m/s^2\; .
\ee
Its value is 7 orders of magnitude smaller than the STEP requirement (\ref{amax}).

Other estimates for the patch parameters are produced by the approach of the previous section using the expression of $|\delta F_{z}|$ averaged over the axial and the angular distances between the patches. As we know, the average of $\mu$ in the formula (\ref{Flin}) is unity, and the average over the axial distance is ${l_0}^{-1}\int^{l_0}_0 d \tilde{z}\left(1-2\tilde{z}^2\right)\,\exp\left[-\tilde{z}^2\right]=\exp(-l_0^2)$ (as before,  $l_0=L/2\Delta z$). Thus the mean angular frequency is found to be
\bea
\bar{\omega}=\sqrt{{\pi^{3/2}}\frac{a_{ax}}{\Delta z}\sin^4\left(\frac{\Delta\vfi}{2}\right)\exp\left[-\left(\frac{L}{2\Delta z}\right)^2\right]}=\nonumber\\
6.2\times 10^{-7}\frac{1}{\sqrt{\Delta z}}\sin^2\left(\frac{\Delta\vfi}{2}\right)\exp\left[-\left(\frac{L}{2\sqrt{2}\Delta z}\right)^2\right]\,rad/s\;,
\label{meanomega}
\eea
with $\Delta z$ on the far right in meters, as usual. As compared to the estimate (\ref{maxomega}), here is an extra power of the sine of $\Delta \vfi$, and, most important, the Gaussian exponent that drops sharply with $L/\Delta z$ growing. Due to this, the conditions (\ref{frcrit}) and (\ref{taucrit}) on the oscillation frequency and runaway time are satisfied for the patches of any sizes, by the estimate (\ref{meanomega}). Something close to this happens if we consider $N$ pairs of patches instead of one. Let us examine the condition (\ref{taucrit}), which is more stringent than (\ref{frcrit}). We can again estimate the overall contribution of $N$ pairs of patches with random voltage signs as $\sqrt {N^2} \delta F_z$, and take the mimimum $\sqrt L(e/2)^{1/4}$ for $\Delta z=L/\sqrt2$. This gives
\be\label{tauforN}
N^{1/2}\,\sin^2\left({\Delta\vfi}/{2}\right)\ll 9.4\,\sqrt L(e/2)^{1/4}=2.7\; ;
\ee
with $L=7\,cm$ cited above. The inequality does not hold always, somewhat limiting the angular patch size to be small enough given $N$. However,  assuming the patches do not overlap, we can use the upper bound (\ref{Nmax}) of $N$ with the above value $\Delta z=L/\sqrt2$ in it. This turns the inequality (\ref{tauforN}) into ${\sin^2\left({\Delta\vfi}/{2}\right)}{\left({\Delta \vfi/2}\right)^{-1/2}}\ll 1.8$. The maximum of the l.h.s here is always less than unity, so the condition (\ref{tauforN}) holds always for non-overlapping patches.

Concluding this section we remind that all the numbers in the estimates were obtained using the patch voltage $V_0=10\,mV$, the TM mass $1\,Kg$, and the TM radius--to--gap ratio $a/d=77$. All the accelerations scale as voltage square, proportional to this ratio, and inversely proportional to the mass; the frequencies and inverse runaway times scale proportional to the voltage, $\sqrt{a/d}$, and the inverse of the square root of the mass.

\subsection{Concluding Remarks. Perspectives of Patch Force Modeling\label{s6.4}}

The main message of the above estimates is that the STEP requirements (\ref{amax}), (\ref{frcrit}) and (\ref{taucrit}) can be met by patches whose sizes and number are appropriately limited; the limits seem not too restrictive and practically achievable. However, those estimates were derived under a number of simplifying assumptions. Perhaps the strongest of them was that all the patches differ at most by the sign of the voltage, having the same voltage magnitude and spot sizes. The results obtained in this paper allow, in fact, for an essentially more realistic patch force modeling, which can be done in the following way.

One represents both patch potentials $V_a(\vfi^{'},z^{'})$ and $V_b(\vfi,z)$ as a superposition of some number, $N_{a,\,b}$, of model patches (\ref{Vpatch}),
\bea
V_\mu(\vfi,z)=\sum\limits_{n=0}^{N_\mu}\cV(\vfi-\vfi_n^\mu, z-z_n^\mu;\,\Delta z_n^\mu,\Delta\vfi_n^\mu)=\nonumber\\
\sum\limits_{n=0}^{N_\mu}V_n^{\,\mu}\, v(\vfi-\vfi_n^\mu, z-z_n^\mu;\,\Delta z_n^\mu,\Delta\vfi_n^\mu),\qquad \mu=a,\;b\; ,\label{distrib}
\eea
with different voltages, sizes, and positions. By the formulas of section \ref{s3} the patch effect forces corresponding to the distributions (\ref{distrib}) can be explicitly calculated, as it is done here for a single patch at each of the boundaries. Being cumbersome, this general calculation is otherwise straightforward, without any new technical difficulties. It leads to the force expression as a quadratic form of the patch voltages $V_n^{\mu}$, with the coefficients depending on all other parameters in a known way.

Having these formulas at hand, one then carries out simulations by specifying parameter sets in various ways and computing the patch forces. One can pick the parameters randomly, and eventually come up with the patch force statistics. One can also use any lab information on the patch distributions, arranging for a semi-random patch sets, as was done, for instance, when simulating magnetic trapped flux distribution on GP-B rotors~\cite{NS}. Such exhaustive analysis can be strongly recommended before the STEP flight. On the other hand, the same general formulas for the transverse forces can be used for fitting control effort data obtained during the experiment, for restoring the voltage patch patterns on the proof masses and bearings. Once the latter are known, the axial forces can be computed, and the systematic experimental error due to them can thus be bounded. Of course, all this is applicable to any experimental set-up with cylindrical geometry.

\begin{acknowledgments}
This work was supported by ICRANet (V.F.) and by KACST through the collaborative agreement with GP-B (A.S.). The authors are grateful to Remo~Ruffini and Francis~Everitt for their permanent interest in and support of this work, and for some valuable remarks. We also thank our colleagues at GP-B and STEP, particularly, Dan DeBra, Sasha Buchman, David Hipkins, John~Mester and Paul Worden for valuable discussions and remarks.
\end{acknowledgments}

\appendix

\section{Calculation of the Transverse and Axial Force for a Single Patch at Each of the Cylinders} \label{A4}

Here we consider one patch at each boundary described by our patch model (\ref{Vpatch}), i.e., for the voltage distributions  (\ref{Onebcp}). The force (\ref{Fint}) due to the interaction between the uniform potential and patches is a linear function of the Fourier coefficients of the boundary patch voltages computed for $k=0$ and $n=0,1,2$. Thus formulas (\ref{VpatchFour}) allow one to obtain ($l_i=\sqrt{2}\Delta z_i,\;i=1,\,2$):
\bea
\label{OneFint}
F_x^{int}=\pi\sqrt{\pi}\frac{\epsilon_0 a}{d^{\,2}}\Vm\left\{
-V_1l_1\frac{1-\lambda_1^2}{2}\cos{\vfi_1}+V_2l_2\frac{1-\lambda_2^2}{2}\cos{\vfi_2}+\right.\nonumber\qquad\qquad\qquad\qquad\qquad\;\;\\
\left.
\frac{x^0}{d}\left[
V_1l_1\left(1-\lambda_1\right)\left(1+\frac{1+\lambda_1}{2}\lambda_1\cos{2\vfi_1}\right)-
V_2l_2\left(1-\lambda_2\right)\left(1+\frac{1+\lambda_2}{2}\lambda_2\cos{2\vfi_2}\right)
\right]+\right.\nonumber\\
\left.
\frac{y^0}{d}\left[
V_1l_1\frac{1-\lambda_1^2}{2}\lambda_1\sin{2\vfi_1}-V_2l_2\frac{1-\lambda_2^2}{2}\lambda_2\sin{2\vfi_2}
\right]
\right\}
\;;\qquad\\
F_y^{int}=\pi\sqrt{\pi}\frac{\epsilon_0 a}{d^{\,2}}\Vm\left\{
-V_1l_1\frac{1-\lambda_1^2}{2}\sin{\vfi_1}+V_2l_2\frac{1-\lambda_2^2}{2}\sin{\vfi_2}+\right.\nonumber\qquad\qquad\qquad\qquad\qquad\;\;\\
\left.
\frac{x^0}{d}\left[
V_1l_1\frac{1-\lambda_1^2}{2}\lambda_1\sin{2\vfi_1}-V_2l_2\frac{1-\lambda_2^2}{2}\lambda_2\sin{2\vfi_2}
\right]+\right.\qquad\nonumber\\
\left.
\frac{y^0}{d}\left[
V_1l_1\left(1-\lambda_1\right)\left(1-\frac{1+\lambda_1}{2}\lambda_1\cos{2\vfi_1}\right)-
V_2l_2\left(1-\lambda_2\right)\left(1-\frac{1+\lambda_2}{2}\lambda_2\cos{2\vfi_2}\right)
\right]
\right\}
\;\nonumber\;.
\eea
The transverse force due to the patch interaction is essentially more cumbersome  to derive, with the additional difficulty of computing some integrals in $k$ and sums over $n$ containing products of Fourier coefficients. We start with $F_x^p$ from the first of the formulas (\ref{Fpz0}). Using the expressions (\ref{VpatchgenFour})and (\ref{Onebcp}),  combined with the equality (\ref{fonk}), we find:
\bea
\label{OneFpx}
F_x^p=-\frac{\pi}{2}\frac{\epsilon_0 a}{d^{\,2}}\int_{-\infty}^{\infty}dk\left\{-\Re\left[
V_1^2 l_1^2 e^{-\frac{k^2l_1^2}{2}}N_1(\lambda_1)e^{\imath \vfi_1}+ V_2^2 l_2^2e^{-\frac{k^2l_2^2}{2}}N_1(\lambda_2)e^{\imath \vfi_2}-\right.\right.\;\;\;\qquad\qquad\quad\;\\
\left.
V_1V_2l_1l_2M_1e^{-\frac{k^2}{4}\left(l_1^2+l_2^2\right)}\left(e^{-ik(z_1-z_2)}+e^{ik(z_1-z_2)}\right)\right]+\nonumber\;\;\\
\frac{x^0}{d}\Re\left[V_1^2 l_1^2 e^{-\frac{k^2l_1^2}{2}}\biggl(N_0(\lambda_1)+N_2(\lambda_1)e^{2\imath \vfi_1}\biggr)+V_2^2 l_2^2e^{-\frac{k^2l_2^2}{2}}\biggl(N_0(\lambda_2)+N_2(\lambda_2)e^{2\imath \vfi_2}\biggr)-\right.\nonumber\;\;\\
\left.
V_1V_2l_1l_2e^{-\frac{k^2}{4}\left(l_1^2+l_2^2\right)}\left(\left(2M_0+M_2\right)e^{-ik(z_1-z_2)}+
M_2e^{ik(z_1-z_2)}\right)\right]+\nonumber\;\\
\frac{y^0}{d}\Im\left[V_1^2 l_1^2 e^{-\frac{k^2l_1^2}{2}}N_2(\lambda_1)e^{2\imath \vfi_1}+V_2^2 l_2^2e^{-\frac{k^2l_2^2}{2}}N_2(\lambda_2)e^{2\imath \vfi_2}-\right.\;\;\nonumber\\
\left.
\left.
V_1V_2l_1l_2M_2e^{-\frac{k^2}{4}\left(l_1^2+l_2^2\right)}\left(e^{-ik(z_1-z_2)}+e^{ik(z_1-z_2)}\right)\right]-\right.\nonumber\\
\left.
kz^0\Im\left[V_1V_2l_1l_2M_1e^{-\frac{k^2}{4}\left(l_1^2+l_2^2\right)}\left(e^{-ik(z_1-z_2)}-e^{ik(z_1-z_2)}\right)
\right]
\right\}\; .\nonumber\;
\eea
Here we have introduced the following notations for the coefficients:
\be
\label{Msdef}
M_q=M_q(\vfi_1,\lambda_1,\vfi_2,\lambda_2)\equiv\frac{1}{{2\pi}}\sum_{n=-\infty}^\infty\,u_n(\lambda_1)e^{-\imath n\vfi_1}\,u^*_{n+q}(\lambda_2)e^{\imath (n+q)\vfi_2}\;;
\ee
\be
N_q(\lambda)\equiv\frac{1}{{2\pi}}\sum_{n=-\infty}^\infty\,u_n(\lambda)\,u^*_{n+q}(\lambda)=M(0,\lambda;0,\lambda)\;;\quad q=0,1,2\; ;
\label{Nsdef}
\ee
$u_n(\lambda)$ defined by the formulas (\ref{uonn}). Coefficients (\ref{Msdef}) are clearly symmetric in the pairs of arguments $\vfi,\lambda$ corresponding to each of the patches, 
$
M_q(\vfi_1,\lambda_1,\vfi_2,\lambda_2)=M_q(\vfi_2,\lambda_2,\vfi_1,\lambda_1) ,
$
which we have already used in the above expressions. Now, formulas (\ref{uonn}) lead to the explicit sums of the series (\ref{Msdef}), since they reduce to geometric progressions:
\bea
\label{Ms}
M_0(\vfi_1,\lambda_1,\vfi_2,\lambda_2)=\frac{\left(1-\lambda_1\right)\left(1-\lambda_2\right)}{4}\left[1+\left(1+\lambda_1\right)\left(1+\lambda_2\right)\frac{\cos(\vfi_1-\vfi_2)-\lambda_1\lambda_2}{2D}\right]\;;\qquad\quad\nonumber\\
M_1(\vfi_1,\lambda_1,\vfi_2,\lambda_2)=\frac{\left(1-\lambda_1\right)\left(1-\lambda_2\right)}{8}\left\{e^{\imath \vfi_1}(1+\lambda_1)\left[1+\frac{\lambda_1}{2}(1+\lambda_2)\frac{e^{\imath(\vfi_1-\vfi_2)}-\lambda_1\lambda_2}{D}\right]+\right.\quad\;\nonumber\\
\left.
e^{\imath \vfi_2}(1+\lambda_2)\left[1+\frac{\lambda_2}{2}(1+\lambda_1)\frac{e^{-\imath(\vfi_1-\vfi_2)}-\lambda_1\lambda_2}{D}\right]
\right\}\;;\qquad\\
M_2(\vfi_1,\lambda_1,\vfi_2,\lambda_2)=\frac{\left(1-\lambda_1\right)\left(1-\lambda_2\right)}{8}\left\{e^{2\imath \vfi_1}(1+\lambda_1)\lambda_1\left[1+\frac{\lambda_1}{2}(1+\lambda_2)\frac{e^{\imath(\vfi_1-\vfi_2)}-\lambda_1\lambda_2}{D}\right]+\right.\nonumber\\
\left.
e^{2\imath \vfi_2}(1+\lambda_2)\lambda_2\left[1+\frac{\lambda_2}{2}(1+\lambda_1)\frac{e^{-\imath(\vfi_1-\vfi_2)}-\lambda_1\lambda_2}{D}\right]+\frac{\left(1+\lambda_1\right)\left(1+\lambda_2\right)}{2}e^{\imath(\vfi_1+\vfi_2)}
\right\}\; ;\qquad\nonumber
\eea
\be
\label{Di}
D=1-2\lambda_1\lambda_2\cos(\vfi_1-\vfi_2)+\left(\lambda_1\lambda_2\right)^2\;.
\ee
By the definition (\ref{Nsdef}), coefficients $N_q(\lambda)$ are found as a particular case of the expressions (\ref{Ms}) when $\lambda_1=\lambda_2=\lambda$ and $\vfi_1=\vfi_2=0$:
\be
\label{Ns}
N_0(\lambda)=\frac{1-\lambda}{8}(3-\lambda);\quad
N_1(\lambda)=\frac{1-\lambda^2}{8}(2-\lambda)\;;\quad
N_2(\lambda)=\frac{1-\lambda^2}{16}(1+4\lambda-3\lambda^2)\;.\qquad
\ee
For the closed form of $F_x^p$, two well known integrals in $k$ are used:
\[
\int_{-\infty}^\infty dk \exp\left[-\frac{k^2\left(l_1^2+l_2^2\right)}{4}\right]e^{\pm\imath k(z_1-z_2)}=\frac{2\sqrt{\pi}}{\sqrt{l_1^2+l_2^2}}\exp\left[-\frac{(z_1-z_2)^2}{l_1^2+l_2^2}\right]\; ;
\]
\[
\int_{-\infty}^\infty dk \exp\left[-\frac{k^2\left(l_1^2+l_2^2\right)}{4}\right]\,k\,e^{\pm\imath k(z_1-z_2)}=
\pm \imath 4\sqrt{\pi}\frac{z_1-z_2}{\left(l_1^2+l_2^2\right)^{{3}/{2}}}\exp\left[-\frac{(z_1-z_2)^2}{l_1^2+l_2^2}\right]\;.
\]
In the case $z_1=z_2$ and $l_1=l_2=l$ the first integral becomes:
$
\int_{-\infty}^\infty dk \exp\left[-\frac{k^2l^2}{2}\right]=\sqrt{2\pi}/l\;.
$
Thanks to all the above results for the series and integrals, we obtain the expression for $F_x^p$, to l. o. in the parameter $r^0/d \ll 1$:
\bea
\label{OneFpxres}
F_x^p=\frac{\pi^{{3}/{2}}}{\sqrt{2}}\frac{\epsilon_0 a}{d^{\,2}}\Biggl\{
\left[
V_1^2 l_1N_1(\lambda_1)\cos\vfi_1+ V_2^2 l_2N_1(\lambda_2)\cos\vfi_2-V_1V_2\bal
\Re\biggl(M_1\biggr)e^{-\tilde z^2}
\right]-\nonumber\\
\frac{x^0}{d}\left[
V_1^2l_1\biggl(N_0(\lambda_1)+N_2(\lambda_1)\cos2\vfi_1\biggr)+V_2^2l_2\biggl(N_0(\lambda_2)+N_2(\lambda_2)\cos2\vfi_2\biggr)-\right.\nonumber\\
\left.
V_1V_2\bal\Re\biggl(M_2+M_0\biggr)e^{-\tilde z^2}
\right]-\label{D9}\\
\frac{y^0}{d}\left[
V_1^2l_1N_2(\lambda_1)\sin2\vfi_1+V_2^2l_2N_2(\lambda_2)\sin2\vfi_2-V_1V_2\bal \Im\biggl(M_2\biggr)e^{-\tilde z^2}
\right]-\nonumber\\
\frac{z^0}{\bal}\,\frac{\bal^{\,2}}{2^{3/2}l_1l_2}\left[
2V_1V_2\bal\Re\biggl(M_1\biggr)\tilde z e^{-\tilde z^2}
\right]
\Biggr\}\;;\nonumber\qquad
\eea
\be
\label{ztilde}
\tilde z=\frac{z_1-z_2}{\sqrt{l_1^2+l_2^2}};\qquad\qquad
\bal=\frac{2^{3/2}l_1l_2}{\sqrt{ l_1^2+l_2^2}}\; .
\ee
Calculation of $F^p_y$ is pretty similar, and its result, to linear order in $r^0/d\ll 1$, is:  
\bea
\label{OneFpyres}
F_y^p=
\frac{\pi^{{3}/{2}}}{\sqrt{2}}\frac{\epsilon_0 a}{d^{\,2}}\Biggl\{
\left[
V_1^2 l_1N_1(\lambda_1)\sin\vfi_1+ V_2^2 l_2N_1(\lambda_2)\sin\vfi_2-V_1V_2\bal
\Im\biggl(M_1\biggr)e^{-\tilde z^2}
\right]-\\
\frac{x^0}{d}\left[
V_1^2l_1N_2(\lambda_1)\sin2\vfi_1+V_2^2l_2N_2(\lambda_2)\sin2\vfi_2-V_1V_2\bal \Im\biggl(M_2\biggr)e^{-\tilde z^2}
\right]-\nonumber\\
\frac{y^0}{d}\left[
V_1^2l_1\biggl(N_0(\lambda_1)-N_2(\lambda_1)\cos2\vfi_1\biggr)+V_2^2l_2\biggl(N_0(\lambda_2)-N_2(\lambda_2)\cos2\vfi_2\biggr)-\right.\nonumber\\
\left.
V_1V_2\bal\Re\biggl(M_0-M_2\biggr)e^{-\tilde z^2}
\right]-\frac{z^0}{\bal}\,\frac{\bal^{\,2}}{2^{3/2}l_1l_2}\left[
2V_1V_2\bal\Im\biggl(M_1\biggr)\tilde z e^{-\tilde z^2}
\right]
\Biggr\}\;.\nonumber\qquad
\eea

Finally, we find a closed--form expression for the axial force. Starting from the last of the formulas (\ref{Fpz0}), we just need to repeat the same steps as with $F_x^p$. The only significant difference is that here we need a slightly different integral,
\[
\int_{-\infty}^\infty dk \exp\left[-\frac{k^2\left(l_1^2+l_2^2\right)}{4}\right]\,k^2\,e^{\pm\imath k(z_1-z_2)}=
\frac{4\sqrt{\pi}}{\left(l_1^2+l_2^2\right)^{3/2}}\left[1-2\frac{(z_1-z_2)^2}{l_1^2+l_2^2}\right]\exp\left[-\frac{(z_1-z_2)^2}{l_1^2+l_2^2}\right]\;.
\]
In this way, using definitions (\ref{ztilde}) and (\ref{Ms}), we arrive at the final expression for $F_z^p$:
\bea
\label{OneFpaxres}
F_z^p=\,\left(\frac{\pi^3}{4}\right)^{1/2}\,\frac{\epsilon_0 a}{d}V_1V_2\frac{\bal^{\,2}}{l_1l_2}e^{-\tilde z^2}\Biggl\{
\tilde z\; \Re\biggl(M_0\biggr)-\quad\;\\
\frac{x^0}{d}\left[\tilde z\; \Re\biggl(M_1\biggr)\right]-
\frac{y^0}{d}\left[\tilde z\; \Im\biggl(M_1\biggr)\right]-
\frac{z^0}{\bal}\,\frac{\bal^{\,2}}{2^{3/2}l_1l_2}\left[\left(1-2\tilde z^{\,2}\right)\;\Re\biggl(M_0\biggr)\right]
\Biggr\}\;.\nonumber\qquad
\eea

The calculation of the patch effect forces for the case when a single patch is placed on each of the cylinders is now completed. The expressions (\ref{OneFpxres}), (\ref{OneFpyres}) and (\ref{OneFpaxres}) are for the patches with different sizes and magnitudes. When those are identical, the results simplify essentially due to $\lambda_1=\lambda_2=\lambda$, $l_1=l_2=l$,  $\left|V_1\right|=\left|V_2\right|$, in particular:
\bea
\label{NewCoeff}
N_0(\lambda)=\left(1-\lambda\right){\cal N}_0\;;\quad 
N_1(\lambda)=\left(1-\lambda\right){\cal N}_1\;;\quad
N_2(\lambda)=\left(1-\lambda\right){\cal N}_2\;;\quad\nonumber\\
M_0(\lambda,\vfi_1,\vfi_2)=\left(1-\lambda\right)
{\cal M}_0\;;\qquad\qquad\qquad\qquad\qquad\\
M_1(\lambda,\vfi_1,\vfi_2)=\left(1-\lambda\right)
{\cal M}_1(\lambda,\vfi_1-\vfi_2)\left[\left(\cos\vfi_1+\cos\vfi_2\right)+\imath\left(\sin\vfi_1+\sin\vfi_2\right)\right]
\;;\nonumber\\
M_2(\lambda,\vfi_1,\vfi_2)=\left(1-\lambda\right)
{\cal M}_2(\lambda,\vfi_1-\vfi_2)\left[\cos\left(\vfi_1+\vfi_2\right)+\imath\sin\left(\vfi_1+\vfi_2\right)\right]\;.\qquad\quad
\nonumber
\eea
The coefficients ${\cal N}_i$ and ${\cal M}_i$ here are given explicitly by the formulas (\ref{samecoeff}).

\vfill\eject
\vskip20mm
\begin{figure}[ht]
\centering
\includegraphics[scale=0.5]{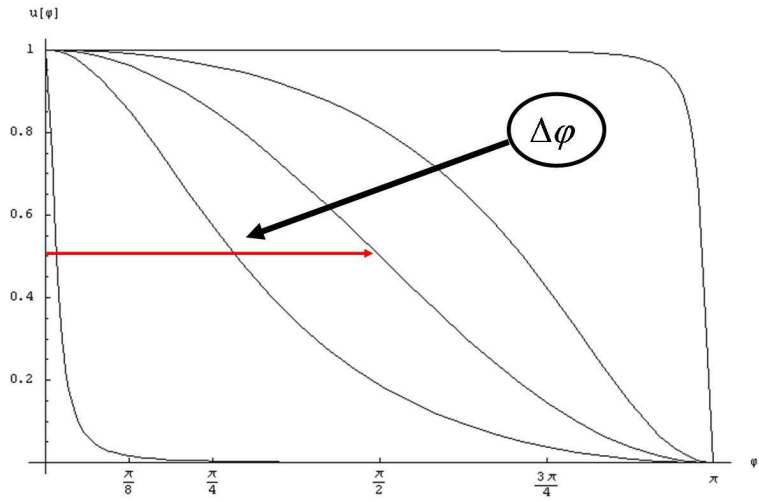}
\caption{{Azimuthal profile of the patch for various 
$\Delta\vfi=1^\circ,\;70^\circ,\;90^\circ,\;110^\circ,\;179^\circ$}}
\label{fig2}
\end{figure}

\vfill\eject

\begin{figure}[ht]
\includegraphics[scale=0.5]{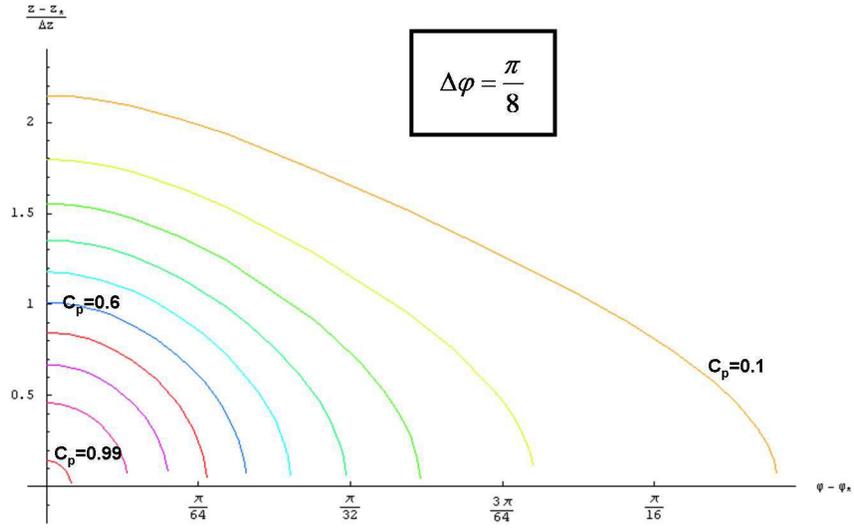}
\caption{{Equipotentials of the patch model for $\Delta\vfi=\pi/8$}}
\label{fig3}
\end{figure}
\vskip20mm
\begin{figure}[ht]
\includegraphics[scale=0.5]{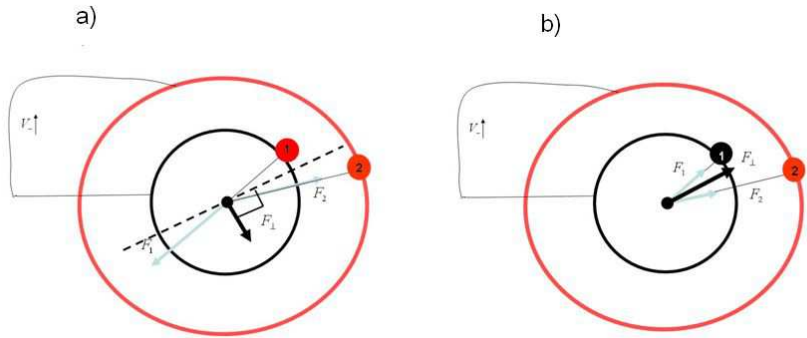}
\caption{{Interaction force due to two patch with a) same sign of the voltages; b) opposite sign of the voltages}}
\label{fig4ab}
\end{figure}

\vfill\eject

\begin{figure}[ht]
\includegraphics[scale=0.5]{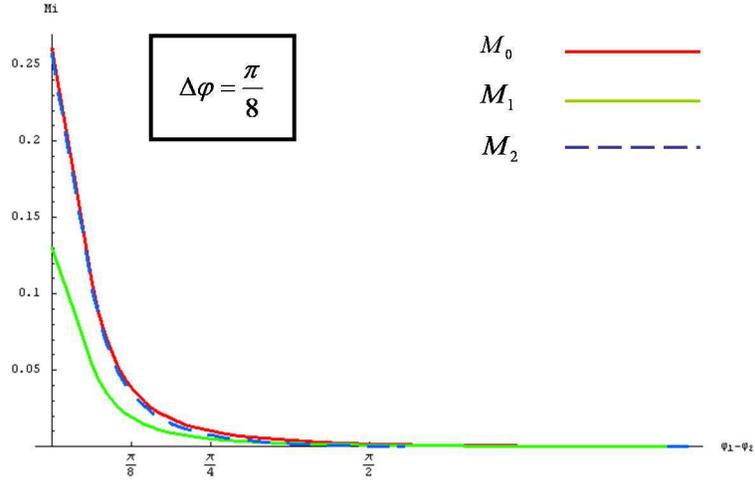}
\caption{{Force coefficients $M_n$  as functions of the angular separation of the patches for $\Delta\vfi=\pi/8$}}
\label{fig5}
\end{figure}
\vskip20mm
\begin{figure}[ht]
\includegraphics[scale=0.5]{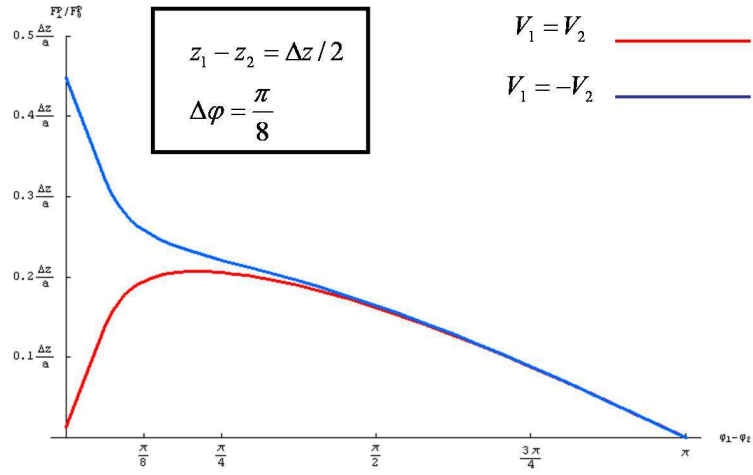}
\caption{Normalized transverse force vs. the angular separation of the patches for $\Delta\vfi=\pi/8$}
\label{fig6}
\end{figure}

\vfill\eject

\begin{figure}[ht]
\includegraphics[scale=0.5]{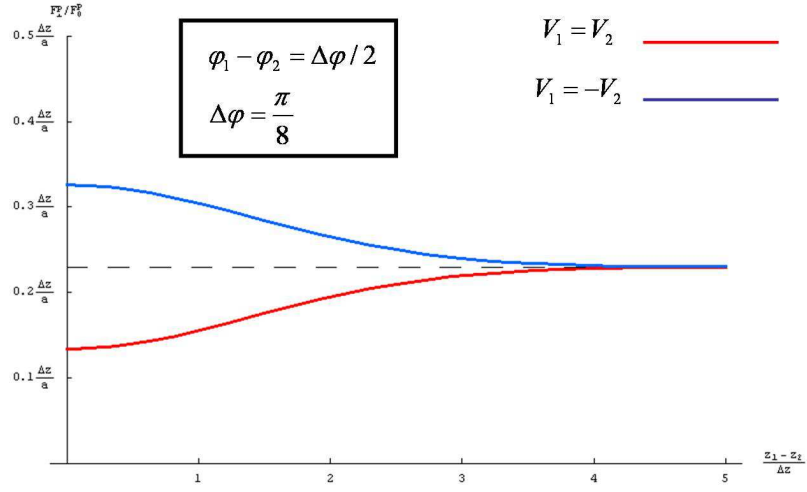}
\caption{{Normalized transverse force vs. the axial separation of the patches}}
\label{fig7}
\end{figure}
\vskip20mm
\begin{figure}[ht]
\includegraphics[scale=0.5]{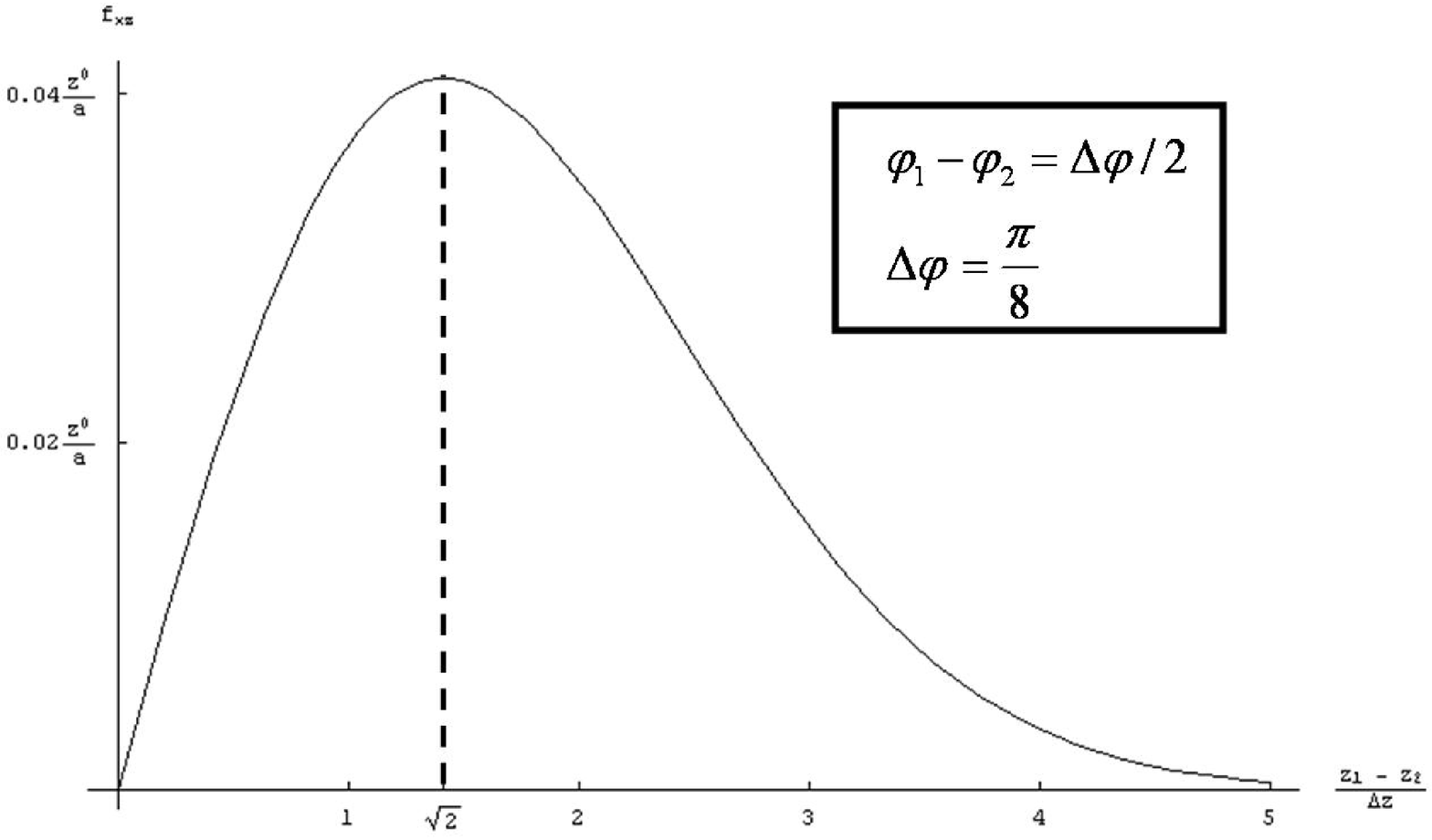}
\caption{Normalized $x$--component of the force due to the axial shift as a function of the axial distance between the patches ($\vfi_2=0$ is taken)}
\label{fig8}
\end{figure}

\vfill\eject

\begin{figure}[ht]
\includegraphics[scale=0.5]{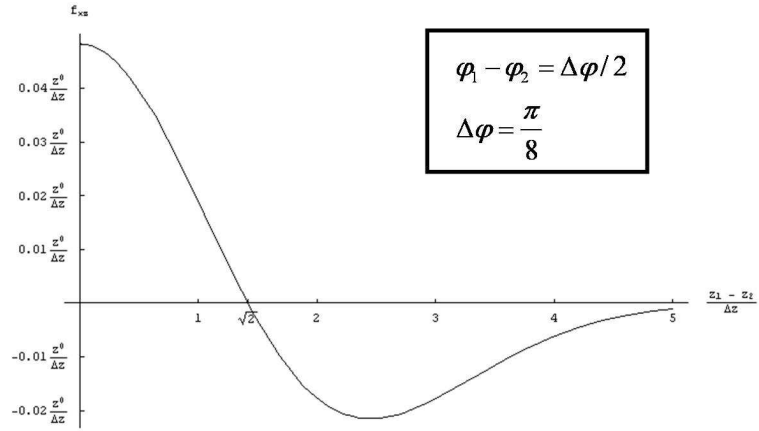}
\caption{Normalized $z$--component of the force due to the axial shift as a function of the axial distance between the patches}
\label{fig9}
\end{figure}
\vskip20mm
\begin{figure}[ht]
\includegraphics[scale=0.5]{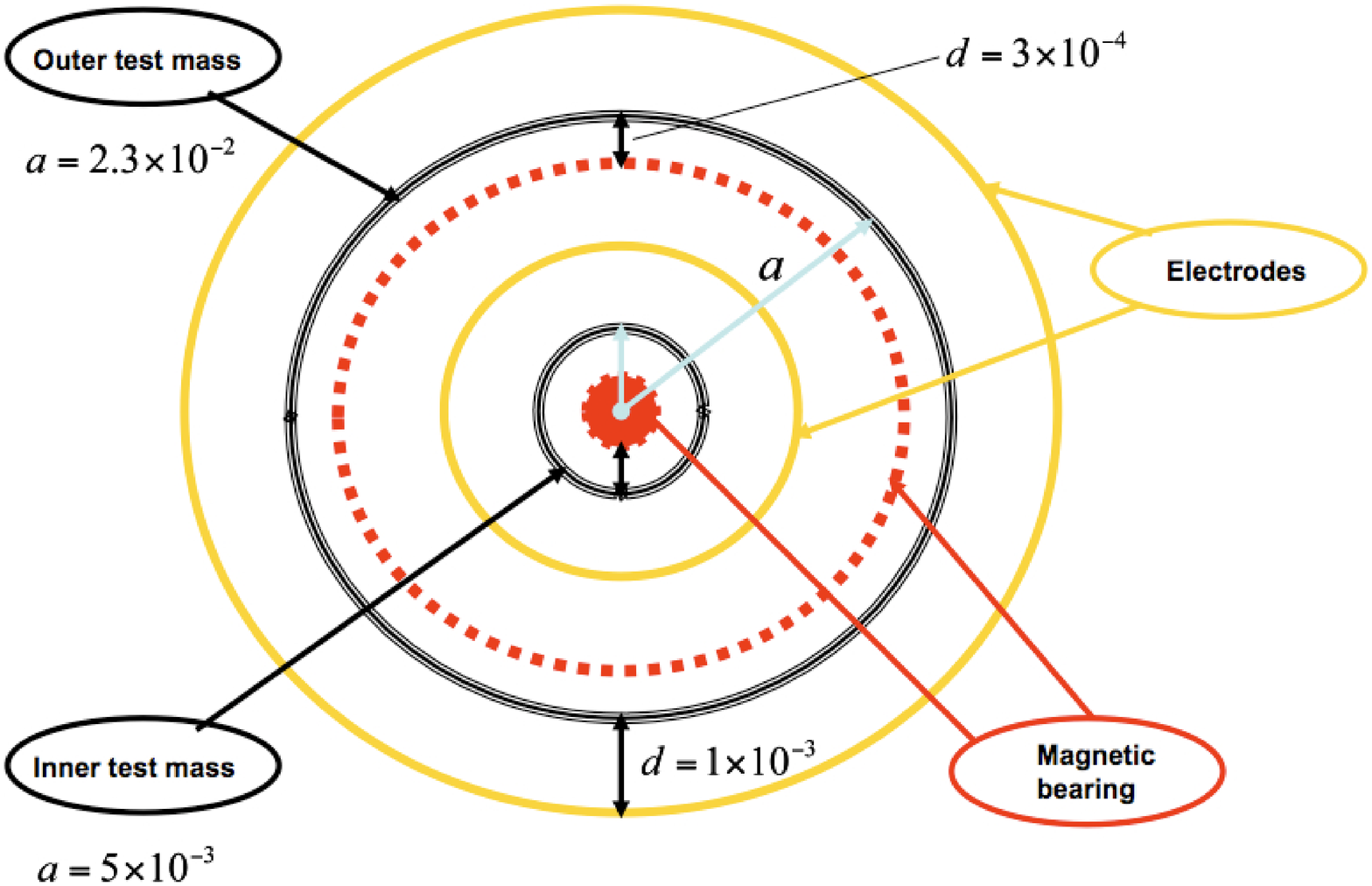}
\caption{Cross-section of the STEP differential accelerometer}
\label{fig10}
\end{figure}


\begin{thebibliography}{99}

\bibitem{VA} Ferroni V., A.S. Silbergleit {\it Electrostatic Patch Effect in Cylindrical Geometry I. Potential and Energy between Slightly Non-Coaxial Cylinders} (submitted to this journ.).
\bibitem{Dar} Darling, T.W. {\it Electric Fields on Metal Surfaces at Low Temperatures}, in: `School of Physics"', University of Melbourne, Parkville, 1989, p.88.
\bibitem{Sp} Speake, C.C. {\it Forces and Force Gradients due to Patch Fields and Contact--Potential Differences}. Class. Quantum. Gravity, {\bf 13}, A291--297, 1996.
\bibitem{LISA} Prince, T., J. Baker, P. Bender, et al. LISA-LIST-RP-436.Version 1.2. A revision of a document originally prepared for the National Research Council {\sl `Beyond Einstein Program Assessment Committee' (BEPAC)} in 2007. 10 March 2009. http://lisa.nasa.gov/
\bibitem{E} Everitt C.W.F., M. Adams, W. Bencze, et al. {\it Gravity Probe B Data Analysis. Status and Potential for Improved Accuracy of Scientific Results}. Space Science Reviews, {\bf 148} (1--4), 53--70 (2009).
\bibitem{H&S} Heifetz M.I., W. Bencze, T. Holmes, A.S. Silbergleit, V. Solomonik. {\it The Gravity Probe B Data Analysis Filtering Approach}. Space Science Reviews, {\bf 148} (1--4), 410--428 (2009).
\bibitem{BT} Buchman S., J. Turneaure,  E. Fei, D. Gill, J.A. Lipa. {\it The Effect of Patch Potentials on the Performance of the Gravity Probe B Gyroscopes}, to be submitted to JAP{\bf Where is it???}.
\bibitem{KetAl} Keiser G.M., M. Adams, W.J. Bencze, et al. {\it Gravity Probe B}. Rivista del Nuovo Cimento  {\bf 32} (11), 555 - 589 (2009).
\bibitem{KKS} Keiser G.M., J. Kolodziejczak, A.S. Silbergleit. {\it Misalignment and Resonance Torques and Their Treatment in the GP-B Data Analysis}. Space Science Reviews,  {\bf 148} (1--4), 383--396 (2009).
\bibitem{PW} Worden Jr., P.W., {\it Almost Exactly Zero: The Equivalence Principle}, Near Zero, 766-782, (1988).
\bibitem{Mes} Mester, J., et al. {\it The STEP mission: principles and baseline design}. Class. Quant. Grav., {\bf 18}, 2475--2486 (2001).
\bibitem{Over} Overduin, J., C.W.F. Everitt, J. Mester, P.W. Worden. {\it The Science Case for STEP}. Adv. in Space Res., {\bf 43}, 1532--1537 (2009).
\bibitem{PWMes} Worden Jr., P.W., J. Mester {\it Satellite Test of the Equivalence Principle Uncertainty Analysis}. Space Science Reviews, {\bf 148} (1--4), 489--499 (2009).
\bibitem{Smy} Smythe, W.R. {\it Static and Dynamic Electricity}, 3rd ed. Hemisphere Publ. Corp.,, New York--Washington--Philadelphia--London, 1989. 
\bibitem{Will} Will C. {\it Progress in Lunar Laser Ranging Tests of Relativistic Gravity}. Phys. Rev. Lett. {\bf 93} 261101 (2004).
\bibitem{NS} Nemenman I.M., A.S. Silbergleit. {\it Explicit Green's function of a boundary value problem for a sphere and trapped flux analysis in Gravity Probe B experiment}. J. Appl. Phys., {\bf 86} (1), 614---24, 1999.

\end{thebibliography}
\end{document}